\documentclass[
reprint,
superscriptaddress,
preprintnumbers,
 amsmath,amssymb,
 aps,
showkeys,
]{revtex4-2}

\usepackage{graphicx}
\usepackage{dcolumn}
\usepackage{bm}
\usepackage{mathptmx}
\usepackage{ulem}
\usepackage{color}
\usepackage{mathtools}
\usepackage{subcaption}
\usepackage[breaklinks=true]{hyperref}

\begin{document}

\preprint{APS/123-QED}

\title{Acoustic radiation force and radiation torque beyond particles: \\
Effects of non-spherical shape and Willis coupling}

\author{Shahrokh Sepehrirahnama}
\email{shahrokh.sepehrirahnama@uts.edu.au} 
\affiliation{%
 Centre for Audio, Acoustics and Vibration, University of Technology Sydney, Sydney, Australia
}

\author{Sebastian Oberst}
\altaffiliation[Also at ]{School of Engineering and Information Technology, University of New South Wales, Canberra, Australia}
\affiliation{%
Centre for Audio, Acoustics and Vibration, University of Technology Sydney, Sydney, Australia
}%

\author{Yan Kei Chiang}%
\altaffiliation[Also at ]{Centre for Audio, Acoustics and Vibration, University of Technology Sydney, Sydney, Australia}
\affiliation{%
 School of Engineering and Information Technology, University of New South Wales, Canberra, Australia
}%

\author{David Powell}
\altaffiliation[Also at ]{Centre for Audio, Acoustics and Vibration, University of Technology Sydney, Sydney, Australia}
\affiliation{%
 School of Engineering and Information Technology, University of New South Wales, Canberra, Australia
}%


\date{\today}

\begin{abstract}
Acoustophoresis deals with the manipulation of sub-wavelength scatterers in an incident acoustic field.
The geometric details of manipulated particles are often neglected by replacing them with equivalent symmetric geometries such as spheres, spheroids, cylinders or disks. It has been demonstrated that geometric asymmetry, represented by Willis coupling terms, can strongly affect the scattering of a small object, hence neglecting these terms may miss important force contributions.
In this work, we present a generalized formalism of acoustic radiation force and radiation torque based on the polarizability tensor, where Willis coupling terms are included to account for geometric asymmetry.
Following Gorkov's approach, the effects of geometric asymmetry are explicitly formulated as additional terms in the radiation force and torque expressions.
By breaking the symmetry of a sphere along one axis using intrusion and protrusion, we characterize the changes in the force and torque in terms of partial components, associated with the direct and Willis Coupling coefficients of the polarizability tensor.
We investigate in detail the cases of standing and travelling plane waves, showing how the \textcolor{black}{equilibrium positions and angles} are shifted by these additional terms.
We show that while the contributions of asymmetry to the force are often negligible for small particles, these terms greatly affect the radiation torque.
Our presented theory, providing a way of calculating radiation force and torque directly from polarizability coefficients, shows that in general it is essential to account for shape of objects undergoing acoustophoretic manipulation, and this may have important implications for applications such as the manipulation of biological cells.
\end{abstract}

\maketitle

\section{\label{sec:intro}Introduction}
\begin{figure}[ht!]
    \centering
    \includegraphics[width=\columnwidth]{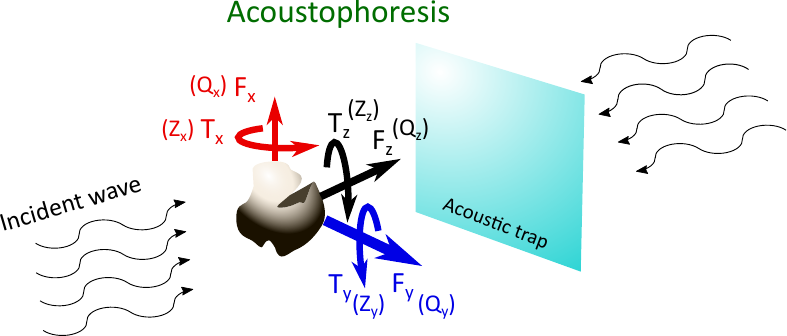}
    \caption{Principle of acoustophoresis with acoustic radiation force and torque exerted on an object with arbitrary shape, for any choice of incident pressure wave and the angle of incidence. The components of the acoustic radiation force and torque and the notation for their normalized contrast factors, $\mathbf{Q}$ and $\mathbf{Z}$, respectively, are illustrated.}
    \label{fig:concept}
\end{figure}

Acoustic radiation forces play a key role in the field of acoustic particle manipulation, also known as acoustophoresis \cite{af_bruus1, af_bruus7, af_Laurell, af_bruus10, af_dual2012, af_wiklund, Lim_2011, Lim_2018}.
Acoustic sorting, separation, levitation and other similar applications have been developed to manipulate particulate phase in a fluid, e.g. to migrate biological cells to certain locations using incident plane waves, by inducing a  radiation force field \cite{af_Laurell, Lim_2011, Laurell2012, garcia2014experimental, Hill_2014, antfolk2015_CTC, DW2015, Lim_2016, Lim_2018, DW2019, memoli2020AcLevMeta}, as schematically shown in Fig.~\ref{fig:concept}.
In the design of such applications, it is customary to treat the objects in the host fluid as spheres or other simple geometries and neglect the details of their shapes.
In most cases, this leads to a design based solely on acoustic radiation force and neglecting the radiation torque.
When a sub-wavelength object is treated as a particle, its dynamic equilibrium and force balance is independent of its shapes and its rotation is neglected.
This assumption can be reasonable for small objects with approximately spherical shape; however, non-spherical objects lacking radial symmetry may be better approximated as rigid bodies than point-like particles.

Acoustic radiation stress consists of radiation pressure, which is related to radiated momentum through scattering fields, and Reynolds stress arising from surface oscillation.
The primary radiation force acting on a scatterer corresponds to the incident-scattering portion of the time-averaged radiation stresses \cite{King_34, Yosioka_55, Gorkov_62, af_bruus7, Shahrokh_2015, silva_Bruus, lopes2016acoustic, Shahrokh_2016, shahrokh2020PRE, Sh2020_agglo}. 
In the case of multiple objects, pairwise secondary radiation forces, also known as acoustic interaction forces, emerge from the scattering-scattering part of the radiation stresses \cite{Doinikov2001_interparticle, silva_Bruus, lopes2016acoustic, Shahrokh_2016, shahrokh2020PRE, Sh2020_agglo}.
The acoustic radiation torque is the resultant moment due to such stresses with respect to the centroid of the object \cite{fan2008torque, zhang2011torque}. 
Acoustic radiation force and torque contribute to the dynamic equilibrium state of the scatterers in combination with other forces such as hydrodynamic drag or gravitational force.
The magnitudes of acoustic radiation force and torque are proportional to the incident energy density and their direction is determined from their normalized value, also referred to as the acoustic contrast factor \cite{Yosioka_55,af_bruus7,shahrokh2020PRE}, as shown in Fig.~\ref{fig:concept}.

\textcolor{black}{
Current theories of acoustic radiation force and torque use either the radiation stresses at the surface of the object \cite{King_34, Yosioka_55, Shahrokh_2015_main, Doinikov1994_JFM, Doinikov1994_Proc}, or convert the surface integral of radiation stresses in the far-field to a volume integral using the Divergence theorem \cite{Gorkov_62, af_bruus7, shahrokh2020PRE}.}
The advantage of the latter approach is that the object can be replaced by a set of acoustic multipole sources to evaluate the volume integral, capturing the essential geometric features and material properties of the object with the simplest possible model. 
Acoustic radiation force and torque have been developed analytically or modelled numerically for objects with spheroidal, cylindrical and disk shapes \cite{Doinikov1994_Proc,fan2008torque,foresti2012ellSpheroid,FBW2015spheroid,mitri2015ellCyl,wei2004cyl,xie2004ARFdisc,garbin2015ARFdisk,Lim_2018b}. 
However, these shapes exhibit a high degree of symmetry (at least axial symmetry), and are not fully representative of three-dimensional objects with arbitrary shape.
\textcolor{black}{Although the numerical approaches based on surface or volume integrals can be applied to objects of arbitrary shape, they do not directly show how the geometric asymmetry impacts on the force and torque. For acoustic scattering from small particles, it has been shown that asymmetry can be accounted for by incorporating Willis coupling terms into the multipole tensor \cite{Alu2017WCorigin, Alu2018maxWC, jordaan2018, anton2019, YK2020arraye}. However, to date there has been no investigation of the role that these terms may play in the acoustic radiation force and torque.}

\textcolor{black}{In this work, we derive a rigorous mathematical formulation of acoustic radiation force and acoustic radiation torque that includes the shape complexity as represented by the Willis coupling terms.
We make use of the far-field approach, also referred to as the Gorkov approach, and show how the particle responds to both the incident pressure and velocity fields.
We present a general formalism, but consider in detail the cases of plane travelling and standing waves. Starting from a symmetric shape, we show how introducing geometric protrusions or instructions controls the Willis coupling, and hence the additional radiation force and torque terms.}

\section{\label{sec:theory}Theory}
\subsection{\label{subsec:WC_theory} Scattering of sub-wavelength objects and polarizability}
The acoustic wave propagation in a lossless fluid is governed by the wave equation , which is expressed in terms of acoustic pressure $p$ as follows,
\begin{align}
    \nabla^2 p = \frac{1}{c_f^2} \partial_{tt} p,
    \label{eq:ac_ge}
\end{align}
where $\nabla^2$ represents the Laplacian operator, $c_f$ denotes the speed of sound in the fluid medium and $\partial_{t} = \frac{\partial}{\partial t}$.
The acoustic density $\rho$ and velocity $\mathbf{v}$ are related to the pressure as follows,
\begin{align}
    p=c_f^2\rho, \quad \partial_t \rho = -\rho_f \nabla\cdot\mathbf{v}, \quad p = -\rho_f\partial_t \phi, \quad \mathbf{v} = \nabla \phi,
    \label{eq:ac_relations}
\end{align}
where $\kappa_f = \rho_f c_f^2$ is the fluid compressibility, \textcolor{black}{$\rho_f$ is the mean fluid density}, and $\phi$ denotes the velocity potential.
Acoustic pressure, density and velocity fields are time-harmonic,
\begin{align}
    p = p(\mathbf{x}) e^{-j\omega t}, \quad \mathbf{v} = \mathbf{v}(\mathbf{x}) e^{-j\omega t}, \quad \rho = \rho(\mathbf{x}) e^{-j\omega t},
    \label{eq:time_dep}
\end{align}
where $t$, $\omega$ and $\mathbf{x}$  denote the time, the angular frequency and the position vector, respectively.

In the Rayleigh limit, the monopole-dipole approximation of the scattering field of a sphere is given as \cite{Gorkov_62, af_bruus7, shahrokh2020PRE}
\begin{align}
\begin{split}
    \phi_s &\approx -\frac{a_s^3}{3\rho_f} f_1 \partial_t  \rho_i \frac{e^{jkr}}{r} - \frac{a_s^3}{2} \nabla\cdot\Big( f_2 \mathbf{v}_i \frac{e^{jkr}}{r} \Big), \\ 
    f_1 &= 1-\frac{\kappa_s}{\kappa_f}, \qquad f_2 = \frac{2\rho_s-2\rho_f}{2\rho_s+\rho_f},
\end{split}
    \label{eq:sc_mo_di_approx}
\end{align}
where $\rho_i$ and $\mathbf{v}_i$ denote the value of incident density and velocity fields, respectively, $a_s$ denotes the sphere radius, $r$ is the radial distance measured from the center, $\rho_s$ and $\kappa_s$ denote the density and compressibility of the scatterer, respectively, and \textcolor{black}{$k=\omega/c_f$} denotes the wave number.
\textcolor{black}{To generalize this approach to arbitrary small objects, including those exhibiting Willis coupling, we employ the method of multipole moments, up to dipole accuracy, and express the scattered pressure as \cite{norris2018, Alu2018maxWC}}
\begin{align}
    \begin{split}
        p_s &\approx -\omega^2 \Omega M G + \omega^2 \pmb{\nabla}\cdot\Big[\Omega\mathbf{D} G \Big],  \quad  G=G\big(kr\big) = \frac{e^{jkr}}{4\pi r}, \\
         \phi_s &= \frac{p_s}{j\rho_f \omega},
    \end{split}
    \label{eq:gen_sc_mo_di_approx}
\end{align}
where $G$ denote the Green's function in 3D domain, and $M$ and $\mathbf{D}$ denote the volumetric monopole and dipole moments, respectively.
\textcolor{black}{The term $a_s^3$ from Eq.~\eqref{eq:sc_mo_di_approx} is generalized to the volume of the object, denoted $\Omega$.}
Furthermore, a characteristic size, denoted by $a$, is required for non-spherical shapes to calculate the Rayleigh index of $ka$.
For this study, we propose
\begin{align}
    a &= 3\frac{\Omega}{\Gamma},
    \label{eq:charac_length}
\end{align}
where $\Gamma$ is the outer surface area.
This allows us to incorporate the volume and surface area as two measures of three-dimensional geometries in our numerical analysis.
A factor of $3$ is included to normalize $a$ to the radius $a_s$ for the case of a spherical object.

The scattering moments are expressed in terms of incident pressure $p_i$ and velocity $\mathbf{v}_i$ fields through the polarization tensor $\pmb{\alpha}$ as follows,
\begin{align}
\begin{split}
    \begin{bmatrix}
    M \\
    \mathbf{D}
    \end{bmatrix}
    =
    \frac{1}{\Omega}
    \begin{pmatrix}
    \alpha_{pp} & \pmb{\alpha}_{pv}^T \\
    \pmb{\alpha}_{vp} & \pmb{\alpha}_{vv}
    \end{pmatrix}
    \begin{bmatrix}
    p_i \\
    \mathbf{v}_i
    \end{bmatrix}, \qquad \pmb{\alpha}_{pv} = \begin{bmatrix}
    \alpha_{pv}^x \\
    \alpha_{pv}^y\\
    \alpha_{pv}^z
    \end{bmatrix},
    \\
    \pmb{\alpha}_{vp} = \begin{bmatrix}
    \alpha_{vp}^x \\
    \alpha_{vp}^y\\
    \alpha_{vp}^z
    \end{bmatrix}, \qquad
  \pmb{\alpha}_{vv} =
  \begin{pmatrix}
    \alpha_{vv}^{xx} & \alpha_{vv}^{xy} & \alpha_{vv}^{xz} \\
    \alpha_{vv}^{yx} & \alpha_{vv}^{yy} & \alpha_{vv}^{yz} \\
    \alpha_{vv}^{zx} & \alpha_{vv}^{zy} & \alpha_{vv}^{zz} 
    \end{pmatrix}
    \label{eq:WC}
\end{split}
\end{align}
where $\pmb{\alpha}_{\rho v}$ and  $\pmb{\alpha}_{v \rho}$ denote the Willis coupling coefficients, $\pmb{\alpha}_{vv}$ denotes direct-dipole polarization tensor, and superscript $T$ denotes the transpose operator.
The Cartesian form of these sub-tensors are given in Eq.~\eqref{eq:WC}.
The entries of $\pmb{\alpha}$ are often normalized as follows \cite{Alu2018maxWC},
\begin{align}
    \begin{bmatrix}
    -\sqrt{3}M \\
    jk\mathbf{D}
    \end{bmatrix}
    &=
    \frac{\overline{\pmb{\alpha}}}{\Omega}
    \begin{bmatrix}
    \frac{1}{\sqrt{3}}p_i\\
    \rho_f c_f \mathbf{v}_i
    \end{bmatrix}, \quad 
    \overline{\pmb{\alpha}} = 
    \begin{pmatrix}
    -3\alpha_{pp} & \frac{-\sqrt{3}}{\rho_f c_f}\pmb{\alpha}_{pv}^T \\
    jk\sqrt{3}\pmb{\alpha}_{vp} & \frac{jk}{\rho_f c_f}\pmb{\alpha}_{vv}
    \end{pmatrix}.
    \label{eq:WC_norm}
\end{align}
By employing the reciprocity principle for Green's function, it has been proven that \cite{Alu2018maxWC}
\begin{align}
    \overline{\pmb{\alpha}} = \overline{\pmb{\alpha}}^{T-} = \begin{pmatrix}
    \overline{\alpha}_{pp} & -\overline{\pmb{\alpha}}_{vp}^T \\
    -\overline{\pmb{\alpha}}_{pv} & \overline{\pmb{\alpha}}_{vv}
    \end{pmatrix}.
    \label{eq:WC_recip}
\end{align}
\textcolor{black}{This yields the relation between the Willis coupling coefficients $\overline{\pmb{\alpha}}_{vp}= -\overline{\pmb{\alpha}}_{pv}$}, which are later required for the formulation of the acoustic radiation force and the acoustic radiation torque.
By substituting Eq.~\eqref{eq:WC} into  Eq.~\eqref{eq:gen_sc_mo_di_approx} and comparing the results with Eq.~\eqref{eq:sc_mo_di_approx}, the monopole and dipole moments for a spherical scatterer become \cite{af_bruus7, shahrokh2020PRE}
\begin{align}
    \begin{bmatrix}
    M \\
    \mathbf{D}
    \end{bmatrix}_{sphere}
    =
    \frac{1}{\Omega}
    \begin{pmatrix}
    \Omega \rho_f \kappa_f f_1 & \mathbf{0}^T \\
    \mathbf{0} & -\frac{3j\Omega}{2\omega}\rho_f f_2 \mathbf{I}
    \end{pmatrix}
    \begin{bmatrix}
    p_i \\
    \mathbf{v}_i
    \end{bmatrix},
    \label{eq:WC_sph}
\end{align}
where $\mathbf{0}$ and $\mathbf{I}$ denote the zero column vector and   the identity matrix of size three, respectively.
The polarizabilty tensor can be expressed as sum of two tensors, as follow,
\begin{align}
    \pmb{\alpha} &= \pmb{\alpha}_{sym} + \pmb{\alpha}_{asym} = \begin{pmatrix}
    \alpha_{pp} & \mathbf{0}^T \\
    \mathbf{0} & \pmb{\alpha}_{vv}
    \end{pmatrix} + 
    \begin{pmatrix}
    0 & \pmb{\alpha}_{pv}^T \\
    \pmb{\alpha}_{vp} & 0\times\mathbf{I}
    \end{pmatrix}
    \label{eq:alpha_decomp}
\end{align}
where $\pmb{\alpha}_{sym}$ denotes the tensor that only includes the direct polarizability coefficients, and $\pmb{\alpha}_{asym}$ denotes the tensor of polarizability arising from pure Willis coupling effect.
We will utilize this decomposition later to characterize the role of direct and Willis coupling coefficients in determining the acoustic radiation force and radiation torque.
Subscripts $sym$ and $asym$ refers to direct and Willis coupling polarizability hereinafter.

Considering the Green's function identity $(\nabla^2+k^2)G=-\delta(r)$, one can write,
\begin{align}
    (\nabla^2-\frac{1}{c_f^2}\partial_{tt})\phi_s \approx -\frac{j\omega}{\rho_f}\Omega M \delta + \frac{j\omega}{\rho_f} \nabla\cdot\Big[\Omega\mathbf{D} \delta \Big],
    \label{eq:dAlambert_op_sc}
\end{align}
where $\delta$ denotes the Dirac delta impulse, and $r$ denotes the distance from the center of the smallest sphere enclosing the scatterer.
We will use Eq.\eqref{eq:dAlambert_op_sc}, which gives the approximation of the scattered field by a set of monopole and dipole sources, to derive the acoustic radiation force and torque.

\subsection{\label{subsec:ARF_theory} Acoustic Radiation Force in the Rayleigh Limit}
Using Gorkov's far-field approach \cite{Gorkov_62, Bruus_2012, af_bruus7, shahrokh2020PRE}, as shown in Supplementary Notes I, the acoustic radiation force acting on a sub-wavelength particle can be expressed as follows,
\begin{align}
    \mathbf{F} &= -\rho_f \int_{\Omega_{\infty}} \Big\langle\mathbf{v}_i \big[\nabla^2-\frac{1}{c_f^2}\partial_{tt}\big]\phi_s \Big\rangle d\Omega,
    \label{eq:f_vol_int}
\end{align}
where $\Omega_{\infty}$ denotes the unbounded volume of the fluid domain, $\langle\cdot\rangle$ denotes the time-averaging operator over one wave period.
For any two harmonically varying fields $\mathcal{F}$ and $\mathcal{G}$, the time-averaged product is $\langle \mathcal{F}\mathcal{G} \rangle = \frac{1}{2}\Re{\big[\mathcal{F}\mathcal{G}^{\ast}\big]}$ with $\Re$ denoting the real part of a complex quantity and $\ast$ denoting the complex conjugation operator.
By substituting Eq.~\eqref{eq:dAlambert_op_sc} into \eqref{eq:f_vol_int} and making use of the properties of the Dirac delta impulse \cite{af_bruus7, shahrokh2020PRE}, the force expression changes to 
\begin{align}
    \mathbf{F} &= \Big\langle j\omega \Omega M \mathbf{v}_i \Big\rangle_{r=0} + \Big\langle j\omega \Omega \mathbf{D}\cdot\pmb{\nabla}\mathbf{v}_i \Big\rangle_{r=0}.
    \label{eq:force_exp_no_int}
\end{align}
Substituting Eq.~\eqref{eq:WC} into \eqref{eq:force_exp_no_int}, the radiation force expression expands in terms of incident fields to
\begin{align}
    \begin{split}
        \mathbf{F} =& \Big\langle j\omega \alpha_{pp} p_i\mathbf{v}_i \Big\rangle_{r=0} + \Big\langle  j\omega \pmb{\alpha}_{p v} \cdot\mathbf{v}_i \mathbf{v}_i \Big\rangle_{r=0}\\
        &+ \Big\langle j \omega p_i\pmb{\alpha}_{vp} \cdot \pmb{\nabla}\mathbf{v}_i \Big\rangle_{r=0} + \Big\langle j\omega \pmb{\alpha}_{vv}\mathbf{v}_i \cdot \pmb{\nabla}\mathbf{v}_i  \Big\rangle_{r=0}.
    \end{split}
    \label{eq:force_exp_expanded}
\end{align}
By using $\pmb{\nabla}p_i=j\omega \rho_f \mathbf{v}_i$, $\langle j\omega\mathcal{F}\mathcal{G}\rangle = -\langle \big(\partial_t\mathcal{F}\big)\mathcal{G}\rangle = \langle \mathcal{F}\big(\partial_t\mathcal{G}\big)\rangle=  -\langle \mathcal{F} j\omega\mathcal{G}\rangle$ and rearranging the terms, the force expression reads
\begin{align}
    \begin{split}
        \mathbf{F} =& -\Big\langle \frac{\alpha_{pp}}{\rho_f} p_i\pmb{\nabla}p_i \Big\rangle_{r=0} + \Big\langle j\omega \pmb{\alpha}_{vv}\mathbf{v}_i \cdot \pmb{\nabla}\mathbf{v}_i  \Big\rangle_{r=0}   \\
        & - \Big\langle \frac{1}{\rho_f}\pmb{\alpha}_{p v}\cdot\mathbf{v}_i \pmb{\nabla}p_i \Big\rangle_{r=0} +  \Big\langle j \omega \pmb{\alpha}_{vp} \cdot \big[p_i \pmb{\nabla}\mathbf{v}_i\big] \Big\rangle_{r=0}.
    \end{split}
    \label{eq:force_exp}
\end{align}
Considering, from \eqref{eq:WC_norm} and Eq.~\eqref{eq:WC_recip} , the relation between Willis coupling coefficients $\pmb{\alpha}_{p v} = j\omega\rho_f\pmb{\alpha}_{vp}$, the force expression simplifies further to
\begin{align}
    \begin{split}
        \mathbf{F} =& -\Big\langle \frac{\alpha_{pp}}{2\rho_f} \pmb{\nabla}\big[p_i^2\big] \Big\rangle_{r=0} + \Big\langle j\omega \pmb{\alpha}_{vv}\mathbf{v}_i \cdot \pmb{\nabla}\mathbf{v}_i  \Big\rangle_{r=0} \\
        &+ \Big\langle \frac{1}{\rho_f}\pmb{\alpha}_{p v}\cdot\big[p_i\pmb{\nabla}\mathbf{v}_i - \mathbf{v}_i\pmb{\nabla}p_i\big] \Big\rangle_{r=0}.
    \end{split}
    \label{eq:force_exp_final}
\end{align}
This general expression of the radiation force holds true for an object of arbitrary shape and of sub-wavelength size.

\subsubsection{\label{sub2sec:f_sph}Acoustic radiation force of a spherical object}
Due to the symmetry, the Willis coupling terms $\pmb{\alpha}_{pv}$ and $\pmb{\alpha}_{vp}$ become zero.
Furthermore, $\pmb{\alpha}_{vv}=\alpha_{vv}\mathbf{I}$
; hence, the second term on the right-hand side of Eq.~\eqref{eq:force_exp_final} changes to
\begin{align}
    \Big\langle j\omega \pmb{\alpha}_{vv}\mathbf{v}_i \cdot \pmb{\nabla}\mathbf{v}_i  \Big\rangle_{r=0} = \Big\langle j\omega\frac{\alpha_{vv}}{2} \pmb{\nabla}\big[v_i^2\big] \Big\rangle_{r=0},
    \label{eq:sph_vGradv}
\end{align}
where $v_i$ is the magnitude of the velocity vector and $v_i^2 = \mathbf{v}_i \cdot \mathbf{v}_i$.
Finally, Eq.~\eqref{eq:force_exp_final} becomes
\begin{align}
    \begin{split}
        \mathbf{F} =& -\Big\langle \frac{\alpha_{pp}}{2\rho_f} \pmb{\nabla}\big[p_i^2\big] \Big\rangle_{r=0} + \Big\langle j\omega\frac{\alpha_{vv}}{2} \pmb{\nabla}\big[v_i^2\big] \Big\rangle_{r=0}.
    \end{split}
    \label{eq:force_sph}
\end{align}
Substituting Eq.\eqref{eq:WC_sph} into Eq.~\eqref{eq:force_sph}, one could derive the Gorkov force potential, which is the basis of radiation force potential theory \cite{Gorkov_62, af_bruus7, Bruus_2012, shahrokh2020PRE}, $\mathbf{F}=-\pmb{\nabla}G$ with
\begin{align}
    G = \frac{4\pi a^3}{3}\Big\langle\frac{f_1}{2}\kappa_f p_i^2 - \frac{3f_2}{4}\rho_f v_i^2 \Big\rangle.
\end{align}

\subsubsection{\label{sub2sec:f_Non_sph}Nonspherical scatterer with rotational and mirror symmetry}

Next, we look at non-spherical geometries which still hold axial or mirror symmetries, such as prolate (elongated) and oblate (flattened) spheroids.
Again, due to \textcolor{black}{mirror} symmetry, the Willis Coupling coefficients become zero.
\textcolor{black}{The three terms on the diagonal of $\pmb{\alpha}_{vv}$ are no longer all equal} and Eq.~\eqref{eq:sph_vGradv} is invalid for an incident wave with an arbitrary 3D wavefront. 
However, for incident plane waves normal to the object's planes of symmetry, Eq.~\eqref{eq:sph_vGradv} can be employed to derive the force potential since the force only acts in the direction of incidence. 
And, for nonspherical scatterers with rotational and mirror symmetry, the force expression becomes
\begin{align}
    \begin{split}
        \mathbf{F} =& -\Big\langle \frac{\alpha_{pp}}{2\rho_f} \pmb{\nabla}\big[p_i^2\big] \Big\rangle_{r=0} + \Big\langle j\omega \pmb{\alpha}_{vv}\mathbf{v}_i \cdot \pmb{\nabla}\mathbf{v}_i  \Big\rangle_{r=0}.
    \end{split}
    \label{eq:f_nonSph_sym}
\end{align}
The expression in Eq.~\eqref{eq:f_nonSph_sym} shows that the Gorkov force potential theory is no longer applicable owing to $\pmb{\alpha}_{vv}\mathbf{v}_i\cdot\pmb{\nabla}\mathbf{v}_i\ne\pmb{\alpha}_{vv}\nabla\big[\mathbf{v}_i\cdot\mathbf{v}_i\big]$, despite the rotational or mirror symmetry of the scatterer.

\subsubsection{\label{sub2sec:f_ge_sph}Acoustic radiation force for objects of arbitrary shape}

A lack of symmetry in the shape of scatterer with respect to the incident field results in Willis coupling.
Since generic shapes can always be found in real-life, it is important to investigate the role of shape complexity and asymmetry. 
For instance red and white blood cells and worm-like bacteria in bio-acoustophoretic applications or bianisotropic metamaterials in acoustic/photonic beam forming, wave manipulation and holography show natural or engineered asymmetries, which are yet to be investigated in the context of acoustic radiation force and radiation torque.
Equation~\eqref{eq:force_exp_final} applies to this general case as long as the characteristic length of the object is within the Rayleigh limit.
Finally, it is evident from Eq.~\eqref{eq:force_exp_final} that Gorkov potential for acoustic radiation force is inapplicable to the general case of objects with arbitrary shapes; hence, the force is required to be calculated directly from Eq.~\eqref{eq:force_exp_final}.

\subsection{\label{sec:ART}Acoustic Radiation Torque}
Acoustic radiation torque $\mathbf{T}$ is obtained from the radiation stresses using the far-field approach, as shown in Supplementary Notes I \cite{Gorkov_62,af_bruus7,maidanik1958torque,fan2008torque,zhang2011torque}, as follows,
\begin{align}
    \mathbf{T} &= -\rho_f \int_{\Omega_{\infty}} \mathbf{x}\times\Big\langle\mathbf{v}_i \big[\nabla^2-\frac{1}{c_f^2}\partial_{tt}\big]\phi_s \Big\rangle d\Omega,
    \label{eq:T_vol_int}
\end{align}
where $\mathbf{x}$ denote the position vector.
Substituting Eq.~\eqref{eq:dAlambert_op_sc} into Eq.~\eqref{eq:T_vol_int} and using $\mathbf{x}\times\mathbf{n} = \mathbf{0}$, the torque expression becomes
\begin{align}
    \mathbf{T} &=  \Big\langle j\omega \Omega\mathbf{D}\times\mathbf{v}_i \Big\rangle_{r=0}.
    \label{eq:T_exp_ge}
\end{align}
This expression is the same as Eq.~(11) in reference \cite{toftul2019CanonicalForm}, in which the radiation force and the radiation torque were derived from their canonical momentum and spin densities.
Finally, using Eq. \eqref{eq:WC}, we obtain the radiation torque for an arbitrarily shaped scatterer, as follows,
\begin{align}
    \mathbf{T} &= \Big\langle j\omega p_i \pmb{\alpha}_{vp}\times\mathbf{v}_i \Big\rangle_{r=0} + \Big\langle j\omega \big[ \pmb{\alpha}_{vv}\mathbf{v}_i \big] \times\mathbf{v}_i \Big\rangle_{r=0}.
    \label{eq:T_exp}
\end{align}
This formulation of radiation torque not only gives the term corresponding to the spin density, which is proportional to $\big\langle \mathbf{v}_i \times\mathbf{v}_i \big\rangle$ \cite{toftul2019CanonicalForm}, but also shows the role of Willis coupling effects distinctively.
Supplementary Notes II outlines the details of how to calculate the polarizability tensor from the scattering of incident standing waves in three dimensions using Boundary Element Method. 

\section{\label{sec:res}Results}

\subsection{\label{sec:SPW}Case of Standing Plane Wave}

The incident pressure, velocity and their derivative fields are expressed as
\begin{align}
    \begin{split}
        p_i &= P_a\cos(kz)e^{-j\omega t},\quad
        \mathbf{v}_i = \frac{jP_a}{\rho_f c_f}\sin(kz)e^{-j\omega t}\mathbf{e}_z,\\
        \pmb{\nabla}p_i &= -P_ak\sin(kz)e^{-j\omega t}\mathbf{e}_z,\quad
        \pmb{\nabla}\mathbf{v}_i = \frac{jP_a k}{\rho_f c_f}\cos(kz)e^{-j\omega t}\mathbf{e}_z\mathbf{e}_z,
    \end{split}
    \label{eq:STW_fields}
\end{align}
Without loss of generality, the propagation direction is assumed to be along the $z$-axis, denoted by $\mathbf{e}_z$.
Substituting Eq.~\eqref{eq:STW_fields} into Eq.~\eqref{eq:force_exp_final} and time averaging, the force and torque acting on an arbitrarily shaped object become
\begin{align}
    \begin{split}
        \mathbf{F}\cdot\mathbf{e}_z =& F_{sym} + F_{asym},\\
        F_{sym} =& \frac{kE_i}{\rho_f}\Big(\frac{\Re{\big[\alpha_{pp}\big]}}{\kappa_f} - k c_f\Im{\big[\pmb{\alpha}_{vv}\mathbf{e}_z\big]}\cdot\mathbf{e}_z\Big) \sin(2kz),\\
        F_{asym} =& \frac{kE_i}{\rho_f} c_f\Im{\big[2\pmb{\alpha}_{pv}\cdot\mathbf{e}_z\big]}\cos(2kz),\\
        \mathbf{T} =& \mathbf{T}_{sym} + \mathbf{T}_{asym},\\
        \mathbf{T}_{sym} =& - \frac{kE_i}{\rho_f} c_f\Im{\big[2\pmb{\alpha}_{vv}\mathbf{e}_z\big]}\times\mathbf{e}_z\sin^2(kz),\\
        \mathbf{T}_{asym} =& \frac{kE_i}{\rho_f}\Big(\frac{1}{\kappa_f}\Re{\big[\pmb{\alpha}_{vp}\big]}\times\mathbf{e}_z\Big) \sin(2kz),
    \end{split}
    \label{eq:f_STW_ge}
\end{align}
where $ E_i = \frac{P_a^2}{4\rho_f c_f^2}$ denotes the acoustic energy density of the incident wave, and $\Im$ denotes the imaginary part of a complex variable.
Subscripts $sym$ and $asym$, previously defined in Eq.~\eqref{eq:alpha_decomp}, refer to the contributions from direct and Willis coupling coefficients, respectively.
$F_{asym}$ and $\mathbf{T}_{asym}$ denote partial force and torque terms that arise from the Willis coupling representation of shape complexity.
The spatial dependence of partial force $F_{asym}$ is $\cos(2kz)$, which gives the stable zero-force location with negative force gradient at $\lambda/8$.
Compared to $\mathbf{F}_{sym}$, which is classically referred to as acoustic radiation force, with $\sin(2kz)$ leading to the prediction of acoustic traps at pressure/velocity nodes under plane standing waves for sub-wavelength spherical and spheroidal particles \cite{Gorkov_62, Yosioka_55, Doinikov1994_Proc, Bruus_2012, FBW2015spheroid, shahrokh2020PRE}, the location of zero net force is shifted by up to $\lambda/8$ along the wave direction.
The effects of geometrical complexity on the primary radiation force is the largest at pressure and velocity nodal planes, where $\sin\big(2kz\big) = 0$.
Furthermore, the actual location of stable zero-force for acoustic traps in a plane standing wave, considering the contribution of Willis coupling, are shifted from the nodal locations as a results of the additional force induced by the Willis Coupling effects.
However, this shift depends on how large the Willis coupling effect is.
This result implies that it is possible to obtain anomalous force and torque fields by engineering the Willis coupling coefficients through shape manipulation.
Finally, it is noted that changing the object symmetry also changes the $\alpha_{pp}$ and $\pmb{\alpha}_{vv}$, since a portion of the scattered energy goes to the Willis coupling effect. 

\subsection{\label{sec:TPW}Case of Travelling Plane Wave}
For a travelling wave in the $z$-direction, the incident pressure, velocity and their derivative fields are expressed as
\begin{align}
    \begin{split}
        p_i &= P_a e^{jkz} e^{-j\omega t},
        \mathbf{v}_i = \frac{P_a}{\rho_f c_f} e^{jkz} e^{-j\omega t}\mathbf{e}_z,\\
        \pmb{\nabla}p_i &= -jk P_a e^{jkz}e^{-j\omega t}\mathbf{e}_z,
        \pmb{\nabla}\mathbf{v}_i = \frac{jP_a k}{\rho_f c_f} e^{jkz} e^{-j\omega t}\mathbf{e}_z\mathbf{e}_z,
    \end{split}
    \label{eq:PTW_fields}
\end{align}
Substituting Eq.\eqref{eq:PTW_fields} into \eqref{eq:force_exp_final}, the force and torque under a plane travelling wave becomes 
\begin{align}
    \begin{split}
         \mathbf{F} =& \frac{2E_i}{\rho_f}\Big(-\Im{\Big[\frac{\alpha_{pp}}{\kappa_f}\Big]}\mathbf{e}_z+k c_f\Re{\Big[\pmb{\alpha}_{vv}\mathbf{e}_z\Big]}\Big),\\
         \mathbf{T} =& -\frac{2kE_i}{\rho_f}\Big(\frac{1}{\kappa_f}\Im{\Big[\pmb{\alpha}_{vp}\Big]}\times\mathbf{e}_z +  c_f\Im{\Big[\pmb{\alpha}_{vv} \mathbf{e}_z\Big]}\times\mathbf{e}_z\Big),
    \end{split}
   \label{eq:f_PTW_z}
\end{align}
This expression shows that the direct contribution of the Willis Coupling terms $\pmb{\alpha}_{pv}$ and $\pmb{\alpha}_{vp}$ to the force under traveling wave is zero.
However, the force generally has also transverse components in $x$- and $y$- direction due to the $\pmb{\alpha}_{vv}\mathbf{e}_z$ term, despite the incident wave's one-dimensional propagation line.

\subsection{\label{sec:Res}Numerical results}
\begin{figure}
    \centering
    \includegraphics[width=0.48\textwidth]{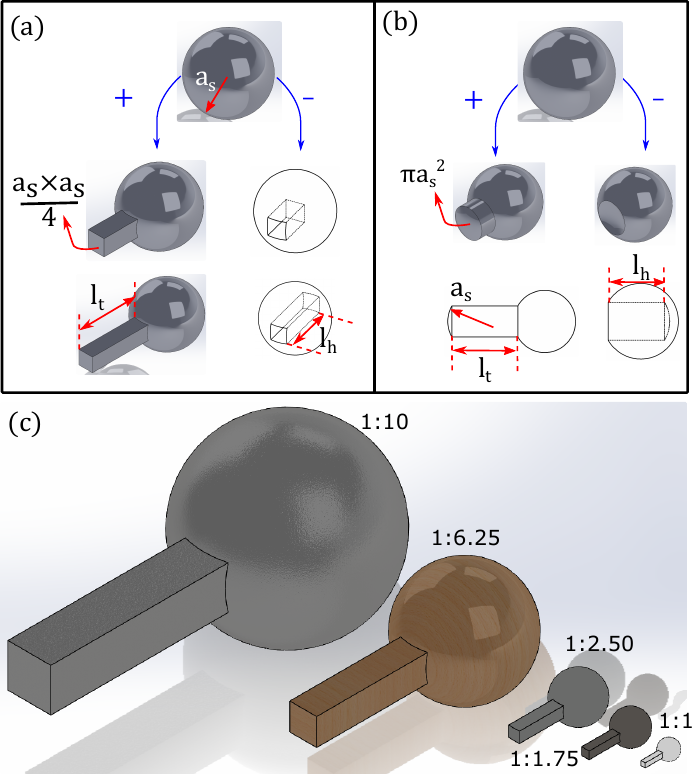}
    \caption{Schematic of non-spherical shapes for the study Willis coupling with engineered anit-symmetry along the $z$-axis by adding, $(+)$, or removing, $(-)$, material at one side.
    Square and circle cross-sections are considered, as shown in panels (a) and (b), respectively, to reduce the spherical symmetry to mirror symmetry in $x-$and $y$-directions, and  axisymmetric respectively.
    Results for the ones with circle cross section are provided in Supplemental Notes III. 
    For the study of size effects, the design with the tail and square cross-section is considered for the given scaling ratios in panel (c), the largest being 10:1 compared to the reference size.
    }
    \label{fig:schematic_sph_tail_hole}
\end{figure}

The shapes of objects are constructed by adding a tail-like attachment (protrusion) to or by creating a hole (intrusion) in a sphere, to engineer a geometrical anti-symmetry in one direction.
The circular tail/hole as shown in Fig.~\ref{fig:schematic_sph_tail_hole} are considered to generate a non-spherical and axisymmetric shape.
The apparent symmetry is further reduced by considering the cross-section of the attachment being a square but having the same edge length as the diameter of the circular one; however, they both exhibit zero Willis coupling in the normal to the length directions due to the mirror symmetry.
The later design was used to investigate the shape effects over a given range of size within the Rayleigh limit, which in practice is reasonably $ka < 0.3$ \cite{sepehrirahnama2015numerical}.

We consider a standing plane wave and assume that objects are sound hard and immovable, allowing us to focus only on the effects of scatterer's exterior shape.
An incident pressure field with $10$ mm wavelength in air with $c_f = 343.140$ $ms^{-1}$ and $\rho_f = 1.204$ $kgm^{-3}$ is considered.
The acoustic radiation force and torque results are normalized, as follows,
\begin{align}
        \mathbf{Q} &=  \frac{\mathbf{F}}{E_i k\Omega}, \qquad
        \mathbf{Z} =   \frac{\mathbf{T}}{E_i \Omega},
    \label{eq:FT_normal}
\end{align}
where $\mathbf{Q}$ and $\mathbf{Z}$  are dimensionless vector quantities, as shown in Fig.~\ref{fig:concept}.
For one dimensional problems such as a sphere in plane waves, these quantities reduce to a scalar value, which has been referred to as force contrast factor an torque contrast factor, respectively, and used to determine the direction of the force and torque \cite{Yosioka_55, Hasegawa_69, Marston, shahrokh2020PRE, silva_Bruus, silva2011_bessel_beam_offaxis, mitri2009Bessel, mitri2015}.

\begin{figure*}
    \begin{subfigure}{0.49\textwidth}
        \centering
        \caption{Real part of non-zero polarizability coefficients}
        \includegraphics[trim={0 0 0 3.2mm}, clip, width=\textwidth]{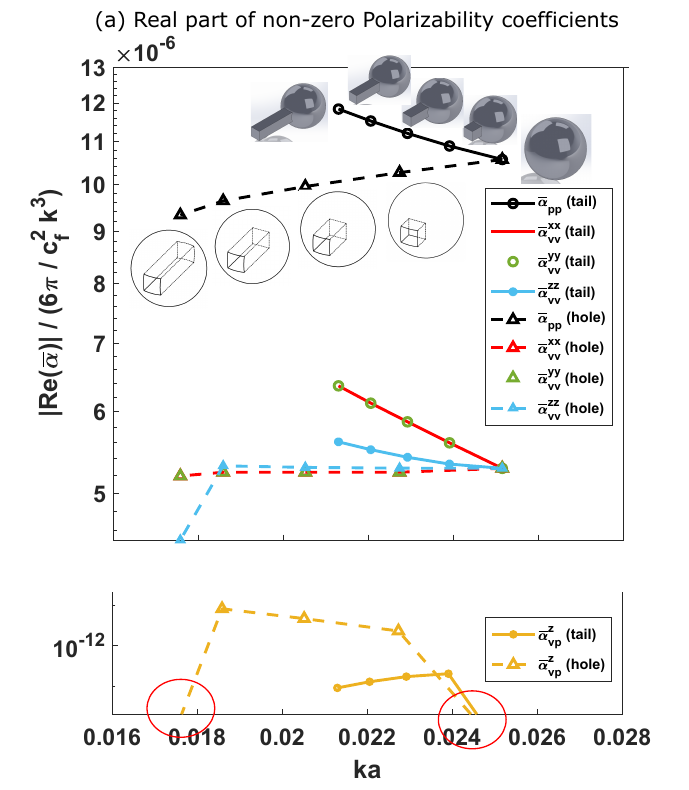}
        \label{fig:alpha_sph_1_a}
    \end{subfigure}
    \begin{subfigure}{0.49\textwidth}
        \centering
        \caption{Imaginary part of non-zero polarizability coefficients}
        \includegraphics[trim={0 0 0 3.6mm}, clip, width=\textwidth]{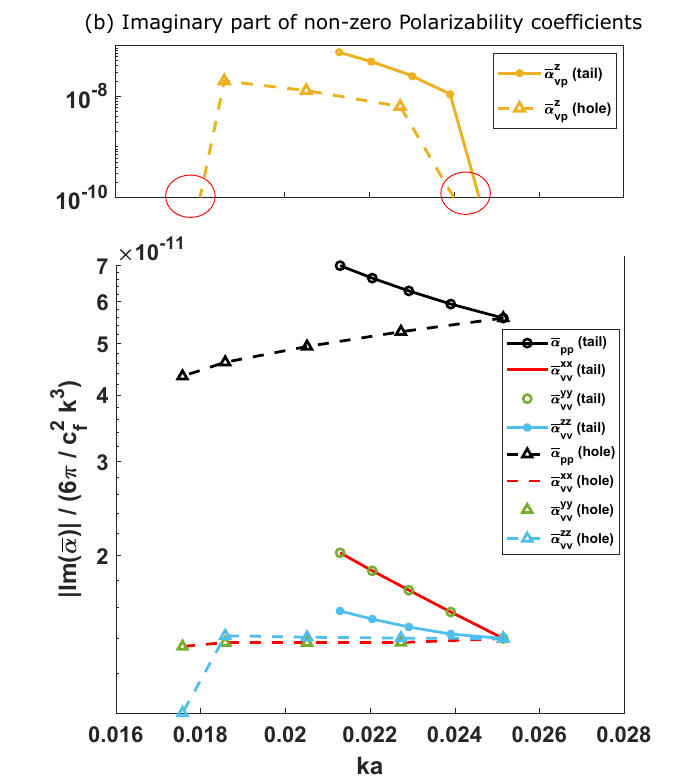}
        \label{fig:alpha_sph_1_b}
    \end{subfigure}
    \caption{Changes of (a) real and (b) imagnary parts of the non-zero polarizability coefficients, given in Eq.~\eqref{eq:WC}, with respect to the deviation from spherical shape by adding a tail or hole with square cross-section.}
    \label{fig:alpha_sph_box}
\end{figure*}

To study the effect of asymmetry in the axial direction $z$, the objects are perturbed from sphere by adding a tail or hole with square cross section.
The non-zero polarizability coefficients are shown in Fig.~\ref{fig:alpha_sph_box}, which indicate that the real parts of Willis coupling coefficients are negligible \textcolor{black}{for the small object sizes considered here}.
For shapes with $z$-symmetry, i.e. the reference sphere and the case with a hole all the way through, both real and imaginary parts of the Willis coupling coefficients become zero, marked by the red circles on the horizontal axis in Fig.~\ref{fig:alpha_sph_box}.
The polarizability coefficients are larger for the cases with the tail, added material volume ($+$), than those with the hole, subtracted material volume ($-$).
Considering Eq.~\eqref{eq:f_STW_ge} which shows the relation between polarizability coefficients and the acoustic radiation force and torque, it is expected that addition of a tail produces larger magnitude of the torque than the cases with a hole.
Finally, a comparison between the cases with square and circular cross-sections is given in Supplemental Notes III, showing that the polarizability coefficients are smaller for the square cross section, for any given tail length or depth hole in the studied range of $0$ to $2a_s$.

\begin{figure*}
    \begin{subfigure}{0.49\textwidth}
        \centering
        \caption{Direct-polarization force $F_{sym}$}
        \includegraphics[trim={0 0 0 2.8mm}, clip, width=\textwidth]{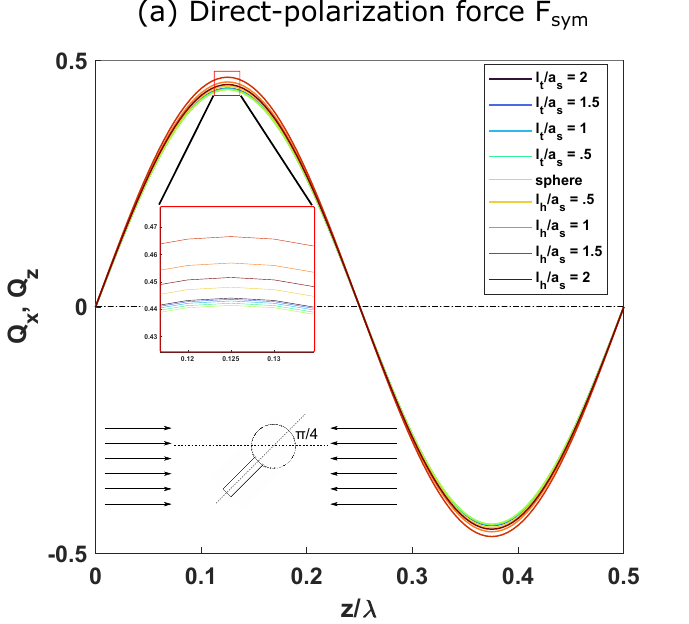}
        \label{fig:box_F_sym}
    \end{subfigure}
    \begin{subfigure}{0.49\textwidth}
        \centering
        \caption{Willis-coupling force $F_{asym}$}
        \includegraphics[trim={0 0 0 2.8mm}, clip, width=\textwidth]{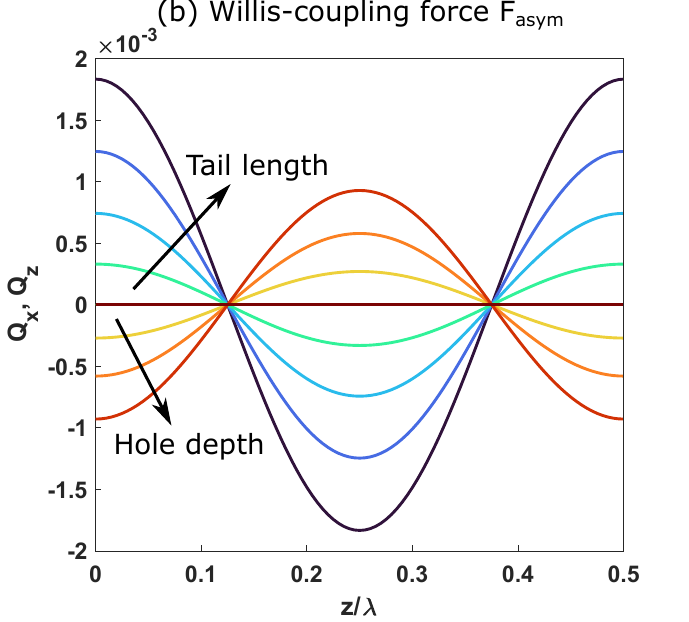}
        \label{fig:box_F_asym}
    \end{subfigure}\\
    \begin{subfigure}{0.49\textwidth}
        \centering
        \caption{Direct-polarization torque $T_{sym}$}
        \includegraphics[trim={0 0 0 3mm}, clip, width=\textwidth]{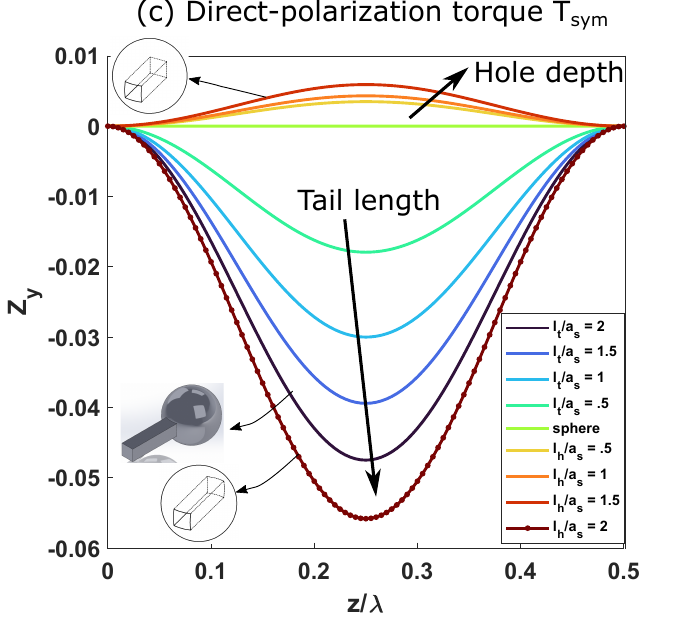}
        \label{fig:box_T_sym}
    \end{subfigure}
    \begin{subfigure}{0.49\textwidth}
        \centering
         \caption{Willis-coupling torque $T_{asym}$}
        \includegraphics[trim={0 0 0 3mm}, clip, width=\textwidth]{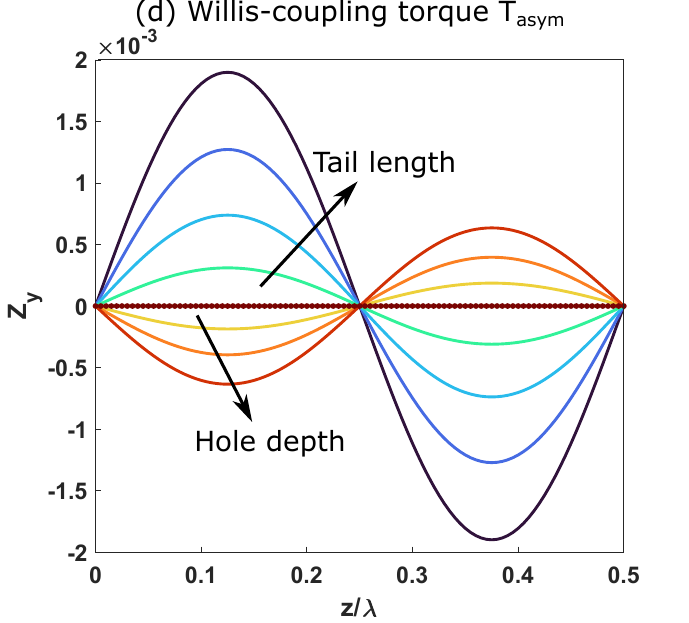}
        \label{fig:box_T_asym}
    \end{subfigure}
    \caption{Normalized force and torque values for the case of a sphere with a rectangular tail/hole, oriented at $\pi/4$ in the $y$-plane. Panels (a) and (c) shows the parts corresponding to the direct polarizability coefficients, while (b) and (d) are those arising from the Willis coupling coefficients.}
    \label{fig:FT_sph_box}
\end{figure*}

To investigate the radiation force and torque further, normalized values $Q_x$, $Q_z$ and $Z_y$ are shown in Fig.~\ref{fig:FT_sph_box}, for the object oriented at $\pi/4$ with respect to the incidence direction, in the $y$-plane.
This means that the radiation force has the same components in the $x$ and $z$ directions, and the torque has only one component in the $y$ direction.
The spatial dependence of the force and torque components are according to Eq.~\eqref{eq:f_STW_ge}.
The normalized values of the partial force $F_{sym}$, due to direct-polarization coefficients, in Fig.~\ref{fig:box_F_sym}, show that the magnitude change is negligible compared to the reference sphere.
However, it was observed that the cases with a tail experiences almost similar forces as the reference sphere.
For those with a hole, the force increases with increasing hole depth, except for the case of through-hole, which shows a sudden force reduction. 
In contrast, the Willis-coupling force $F_{asym}$ varies more significantly as the tail length or the depth hole increases.
The only exception is the case of through hole, $l_h/a_s = 2$, which gives zero Willis-coupling force due to $z$-symmetry.
Nonetheless, the Willis-coupling force $F_{asym}$ is at least two orders of magnitude smaller than the direct-polarization force $F_{sym}$; hence, it could be neglected for estimating the radiation force for practical applications.

The results of torque contrast factor $Z_y$, in Fig.\ref{fig:box_T_sym}-\ref{fig:box_T_asym}, shows that radiation torque is more sensitive to the deviation from spherical shape.
The changes of $T_{sym}$ are larger for the case with the tail, experiencing a negative torque that aligns the tail with the direction of the incident wave vector.
Those with a hole are subjected to a positive torque that tends to align the hole in the normal to the wave vector direction.
The case with a through hole is an exception as it experiences a large negative torque, similar to the cases with a tail.
These results imply that the partial torque $T_{sym}$, due to direct polarization effect, tends to bring the object to the orientation with the smallest cross-section normal to the incident wave direction.
Similar relation between object orientation and the radiation torque was observed for prolate and oblate spheroids \cite{FBW2015spheroid}.

However, as shown in Fig.~\ref{fig:box_T_asym}, these non-spherical objects are also subjected to the Willis-coupling torque $T_{asym}$, as expressed in Eq.~\eqref{eq:f_STW_ge}.
Unlike the large difference between partial radiation forces, this partial torque is just smaller than the $T_{sym}$ by less than an order of magnitude.
It was found that $T_{asym}$ opposes and reinforces the $T_{sym}$ before and after the pressure node at $z/\lambda = 0.25$, respectively.
Therefore, the orientation of the objects under the action of the radiation torque needs to be determined by accounting for both components.
These observations are particularly of interest as it captures not only the contribution of Willis coupling, but also the significance of calculating radiation torques in capturing the essence of shape complexity, even for sub-wavelength objects with $ka\ll 1$.

\begin{figure*}
    \begin{subfigure}{0.49\textwidth}
        \centering
        \caption{Real part of non-zero polarizability coefficients}
        \includegraphics[trim={0 0 0 5mm}, clip, width=\textwidth]{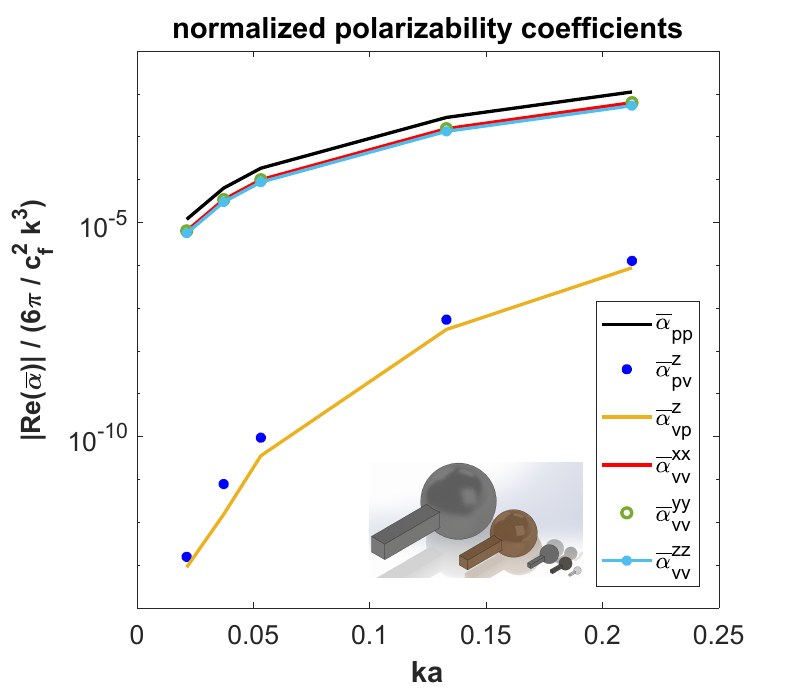}
        \label{fig:alpha_sph_size_a}
    \end{subfigure}
    \begin{subfigure}{0.49\textwidth}
        \centering
        \caption{Imaginary part of non-zero polarizability coefficients}
        \includegraphics[trim={0 0 0 5mm}, clip, width=\textwidth]{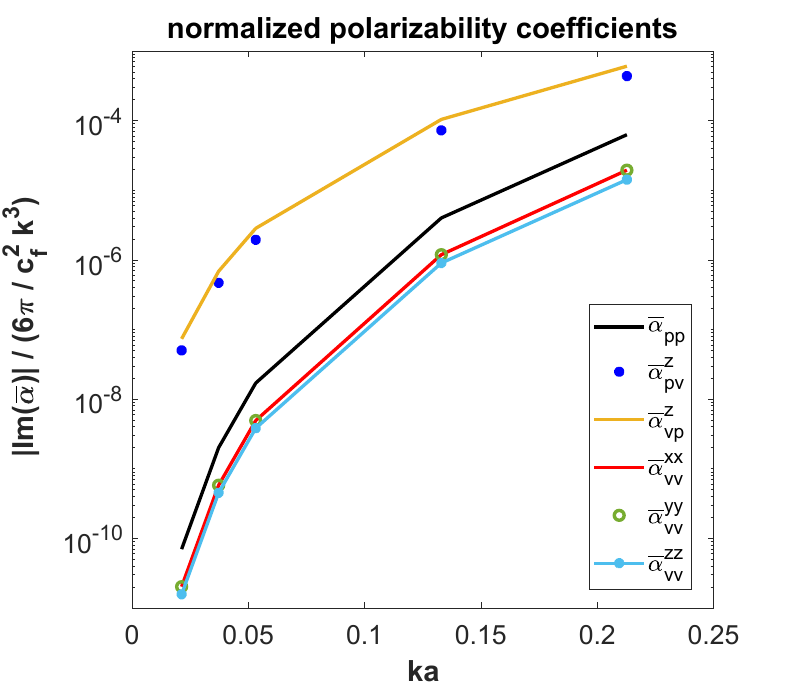}
        \label{fig:alpha_sph_size_b}
    \end{subfigure}
    \caption{Variation of (a) real and (b) imagnary parts of the non-zero polarizability coefficients in terms of size for the case of adding a tail with square cross section.}
    \label{fig:alpha_sph_size}
\end{figure*}

The changes of polarizability response in terms of size were investigated for the case of a tail with square cross section and $l_t/a_s = 2$.
A scaling ratio from 1:1 to 1:10 was applied, indicating an increase of size by one order of magnitude, as shown by the $ka$ values in Fig.~\ref{fig:alpha_sph_size}.
It was found that both real and imaginary parts of the  non-zero polarization coefficients increases  with increasing size ratio.
Real part of Willis-coupling coefficients $\alpha_{pv}^z$ increases by at least seven orders of magnitude, while the direct polarizability coefficients grow by three.
For the imaginary parts in Fig.~\ref{fig:alpha_sph_size_b}, the growth of Willis-coupling is around four orders of magnitude while it is six orders for the direct polarizability coefficients.
These changes of polarizability coefficients influences the radiation force and torque, as can be seen in Eq.~\eqref{eq:f_STW_ge}.
Moreover, our results of $\alpha_{pv}^z$ and $\alpha_{vp}^z$ indicate that the reciprocity principle for the acoustic polarization, as expressed in Eq.~\eqref{eq:WC_recip}, was satisfied within the computational margin of accuracy over the given size range.

\begin{figure}
    \begin{subfigure}{0.48\textwidth}
        \centering
        \caption{Radiation force components $F_{sym}$ and $F_{asym}$  }
        \includegraphics[trim={0 0 0 5mm}, clip, width=\textwidth]{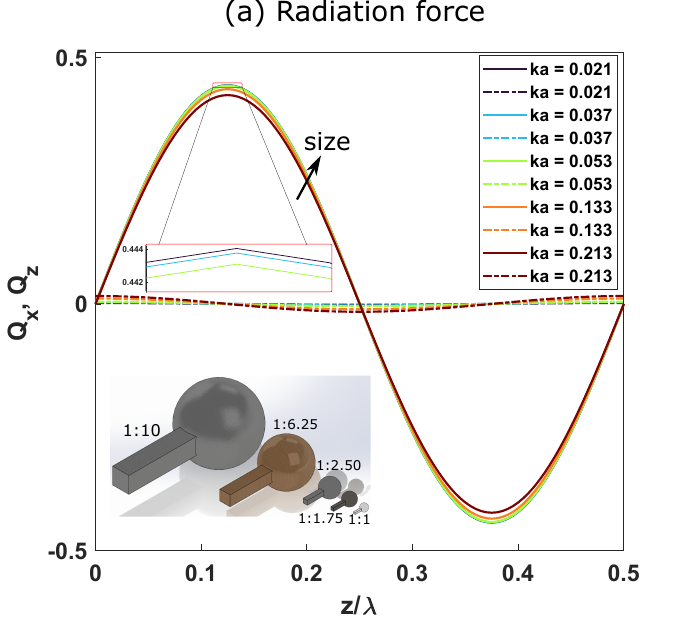}
        \label{fig:F_box_size}
    \end{subfigure}
    \begin{subfigure}{0.48\textwidth}
        \centering
        \caption{Radiation torque components $T_{sym}$ and $T_{asym}$}
        \includegraphics[trim={0 0 0 5mm}, clip, width=\textwidth]{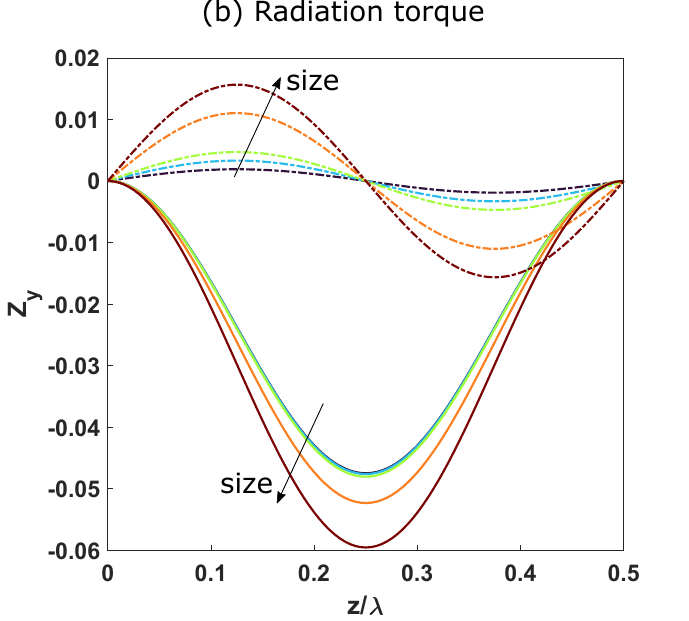}
        \label{fig:T_box_size}
    \end{subfigure}
    \caption{Variation of normalized radiation force and radiation torque in terms of size for the case of a sphere with a rectangular tail, oriented at $\pi/4$ in the $y$-plane. Solid lines indicate $F_{sym}$ and $T_{sym}$ and the other parts are shown by dashed lines.}
    \label{fig:FT_box_size}
\end{figure}

\textcolor{black}{Inspecting Eq.~\eqref{eq:f_STW_ge}, we expect the changes of polarizability coefficients with size to influence the radiation force and torque, and our results are shown in Fig.~\ref{fig:FT_box_size}.}
Although the Willis coupling increased for larger objects, its contribution to the radiation force is still negligible.
The radiation force is less sensitive to size increase as the spread of forces is rather narrow, indicating almost the same magnitude for the given size range.
$T_{asym}$ increased to one third of $T_{sym}$, which implies a greater contribution from the Willis coupling for larger objects.
Both parts of the radiation torque increase faster for $ka > 0.05$, which shows a nonlinear change with respect to the size factor.
These results demonstrate that the effects of size and shape on the radiation torque are more prominent than the radiation force.

\section{\label{sec:dis}Discussion}
In the presented formulation, the assumption of a sub-wavelength scatterer, $ka < 1$, led to the monopole-dipole approximation of the scattering field.
For $ka > 1$, a more accurate approximation including the quadrupole and higher order multipole moments is required to obtain the analytical expressions of the acoustic radiation force and radiation torque.
This could be achieved by incorporating the Willis coupling factors into the partial-wave expansion series and adjusting the scattering coefficients accordingly.

We also observed that the normalized force, also called contrast factor, shows far less variation across the presented range of $ka$ values (Rayleigh index), when $k\Omega$ is used instead of the choice of $\pi a^2$.
The difference is because $k\Omega\propto \pi a^2 \times ka$.
In previous studies mainly focused on spheres or spheroids, the extra $ka$ factor was always applied to the contrast factor, leading to a size dependence which bears no helpful information \cite{Yosioka_55, Hasegawa_69, lopes2016acoustic, mitri2015, shahrokh2020PRE}.
Therefore, it is concluded that the normalisation of radiation force and torque with the respect to volume, as expressed in Eq.~\eqref{eq:FT_normal}, better indicates the dependence on the shape and size features when it comes to non-spherical objects.

The assumptions of sound-hard immovable scatterers were made to focus this study on the effects of shape and size.
\textcolor{black}{However, including the effects of material properties and wave refraction is straightforward, since these can be readily incorporated into the multipole moments of a scatterer.}
Our choice of numerical approach for solving scattering problem was Boundary Element Method, which was combined by the analytical multipole translation and rotation to obtain the multipole moments from the four reference cases of plane standing wave.
The alternative approach is to use Finite Element Method, which is available in commercial software packages, and use the polarizability retrieval techniques \cite{norris2018,jordaan2018}, based on surface integral of scattering pressure with appropriate spherical harmonics as weight functions.
Nevertheless, it is noted that the discretisation of the scatterer's surface for either of these methods requires extra care to ensure a uniform distribution of element size and aspect ratio across the surface.
Our results were obtained after a numerical convergence test to achieve the optimal element size.
The BEM scattering results was verified by comparing against FEM simulation of pressure on a fictitious spherical surface at $4\lambda$ distance, with relative difference of less than $1\%$, for non-spherical objects.

The advantage of using the presented novel formulation is its ability to calculate acoustic radiation force and acoustic radiation torque straightaway from the polarizability tensor, as a measure of acoustic transfer function and being independent of the incident wave field.
If non-planar acoustic waves such as Bessel or Gaussian beams are of interest, one could simply calculate the the radiation force and torque from the incident field values at the centroid of the object.
Moreover, the effect of the angle of incidence is easily included by rotation of the polarizability tensor.
For sub-wavelength objects, our explicit force and torque expressions can be implemented into commercial software packages such as COMSOL for multi-physics simulations of acoustofluidic processes.  
Our results show for the first time that geometrical complexities have no effects on the radiation force that is induced by a plane traveling wave.
However, in the case of two counter-propagating waves, which gives the plane standing wave scenario, the radiation force and radiation torque will be influenced by such shape effects.
This is an indication of the non-linear nature of radiation force and radiation torque, for which the  wave superposition principle becomes inapplicable.

\section{\label{sec:con}Conclusion}
The theory of acoustic radiation force and torque was revisited to incorporate the Willis coupling arising from the shape complexities of an arbitrary object in incident acoustic fields.
The mathematical expression of the acoustic radiation force and radiation torque were provided in terms of polarizability coefficients, applicable for any choice of incident pressure field.
As examples, we derived these expressions for plane standing and travelling waves to characterize the impact of Willis coupling and changes to the polarizability tensor as compared to the case of simple sphere.
We found that radiation torque is more sensitive to the shape effects than radiation force.
As the size factor $ka$ increases, the contribution of the Willis coupling effect becomes larger for the radiation torque, while it remains negligible to the radiation force.
It is concluded that knowing the acoustic radiation torque becomes imperative when the geometry of the object is accounted for even within the Rayleigh limit.
Furthermore, there will be locations in a standing wave with stable angular balance and zero radiation torque.
The presented work outlined the significance of the acoustic radiation force and  radiation torque in the analysis and design of acoustophoresis processes involving non-spherical objects.

\section*{\label{sec:ack}Acknowledgement}
This research is supported by Australian Research
Council Discovery Project DP200101708.

\bibliography{ARC_project_source}

\begin{thebibliography}{53}%
\makeatletter
\providecommand \@ifxundefined [1]{%
 \@ifx{#1\undefined}
}%
\providecommand \@ifnum [1]{%
 \ifnum #1\expandafter \@firstoftwo
 \else \expandafter \@secondoftwo
 \fi
}%
\providecommand \@ifx [1]{%
 \ifx #1\expandafter \@firstoftwo
 \else \expandafter \@secondoftwo
 \fi
}%
\providecommand \natexlab [1]{#1}%
\providecommand \enquote  [1]{``#1''}%
\providecommand \bibnamefont  [1]{#1}%
\providecommand \bibfnamefont [1]{#1}%
\providecommand \citenamefont [1]{#1}%
\providecommand \href@noop [0]{\@secondoftwo}%
\providecommand \href [0]{\begingroup \@sanitize@url \@href}%
\providecommand \@href[1]{\@@startlink{#1}\@@href}%
\providecommand \@@href[1]{\endgroup#1\@@endlink}%
\providecommand \@sanitize@url [0]{\catcode `\\12\catcode `\$12\catcode
  `\&12\catcode `\#12\catcode `\^12\catcode `\_12\catcode `\%12\relax}%
\providecommand \@@startlink[1]{}%
\providecommand \@@endlink[0]{}%
\providecommand \url  [0]{\begingroup\@sanitize@url \@url }%
\providecommand \@url [1]{\endgroup\@href {#1}{\urlprefix }}%
\providecommand \urlprefix  [0]{URL }%
\providecommand \Eprint [0]{\href }%
\providecommand \doibase [0]{https://doi.org/}%
\providecommand \selectlanguage [0]{\@gobble}%
\providecommand \bibinfo  [0]{\@secondoftwo}%
\providecommand \bibfield  [0]{\@secondoftwo}%
\providecommand \translation [1]{[#1]}%
\providecommand \BibitemOpen [0]{}%
\providecommand \bibitemStop [0]{}%
\providecommand \bibitemNoStop [0]{.\EOS\space}%
\providecommand \EOS [0]{\spacefactor3000\relax}%
\providecommand \BibitemShut  [1]{\csname bibitem#1\endcsname}%
\let\auto@bib@innerbib\@empty
\bibitem [{\citenamefont {Bruus}(2011)}]{af_bruus1}%
  \BibitemOpen
  \bibfield  {author} {\bibinfo {author} {\bibfnamefont {H.}~\bibnamefont
  {Bruus}},\ }\bibfield  {title} {\bibinfo {title} {Acoustofluidics 1:
  Governing equations in microfluidics},\ }\href@noop {} {\bibfield  {journal}
  {\bibinfo  {journal} {Lab on a Chip}\ }\textbf {\bibinfo {volume} {11}},\
  \bibinfo {pages} {3742} (\bibinfo {year} {2011})}\BibitemShut {NoStop}%
\bibitem [{\citenamefont {Bruus}(2012{\natexlab{a}})}]{af_bruus7}%
  \BibitemOpen
  \bibfield  {author} {\bibinfo {author} {\bibfnamefont {H.}~\bibnamefont
  {Bruus}},\ }\bibfield  {title} {\bibinfo {title} {Acoustofluidics 7: The
  acoustic radiation force on small particles},\ }\href@noop {} {\bibfield
  {journal} {\bibinfo  {journal} {Lab on a Chip}\ }\textbf {\bibinfo {volume}
  {12}},\ \bibinfo {pages} {1014} (\bibinfo {year}
  {2012}{\natexlab{a}})}\BibitemShut {NoStop}%
\bibitem [{\citenamefont {Lenshof}\ \emph {et~al.}(2012)\citenamefont
  {Lenshof}, \citenamefont {Magnusson},\ and\ \citenamefont
  {Laurell}}]{af_Laurell}%
  \BibitemOpen
  \bibfield  {author} {\bibinfo {author} {\bibfnamefont {A.}~\bibnamefont
  {Lenshof}}, \bibinfo {author} {\bibfnamefont {C.}~\bibnamefont {Magnusson}},\
  and\ \bibinfo {author} {\bibfnamefont {T.}~\bibnamefont {Laurell}},\
  }\bibfield  {title} {\bibinfo {title} {Acoustofluidics 8: Applications of
  acoustophoresis in continuous flow microsystems},\ }\href@noop {} {\bibfield
  {journal} {\bibinfo  {journal} {Lab on a Chip}\ }\textbf {\bibinfo {volume}
  {12}},\ \bibinfo {pages} {1210} (\bibinfo {year} {2012})}\BibitemShut
  {NoStop}%
\bibitem [{\citenamefont {Bruus}(2012{\natexlab{b}})}]{af_bruus10}%
  \BibitemOpen
  \bibfield  {author} {\bibinfo {author} {\bibfnamefont {H.}~\bibnamefont
  {Bruus}},\ }\bibfield  {title} {\bibinfo {title} {Acoustofluidics 10: scaling
  laws in acoustophoresis},\ }\href@noop {} {\bibfield  {journal} {\bibinfo
  {journal} {Lab on a Chip}\ }\textbf {\bibinfo {volume} {12}},\ \bibinfo
  {pages} {1578} (\bibinfo {year} {2012}{\natexlab{b}})}\BibitemShut {NoStop}%
\bibitem [{\citenamefont {Dual}\ \emph {et~al.}(2012)\citenamefont {Dual},
  \citenamefont {Hahn}, \citenamefont {Leibacher}, \citenamefont {M{\"o}ller},
  \citenamefont {Schwarz},\ and\ \citenamefont {Wang}}]{af_dual2012}%
  \BibitemOpen
  \bibfield  {author} {\bibinfo {author} {\bibfnamefont {J.}~\bibnamefont
  {Dual}}, \bibinfo {author} {\bibfnamefont {P.}~\bibnamefont {Hahn}}, \bibinfo
  {author} {\bibfnamefont {I.}~\bibnamefont {Leibacher}}, \bibinfo {author}
  {\bibfnamefont {D.}~\bibnamefont {M{\"o}ller}}, \bibinfo {author}
  {\bibfnamefont {T.}~\bibnamefont {Schwarz}},\ and\ \bibinfo {author}
  {\bibfnamefont {J.}~\bibnamefont {Wang}},\ }\bibfield  {title} {\bibinfo
  {title} {Acoustofluidics 19: Ultrasonic microrobotics in cavities: devices
  and numerical simulation},\ }\href@noop {} {\bibfield  {journal} {\bibinfo
  {journal} {Lab on a Chip}\ }\textbf {\bibinfo {volume} {12}},\ \bibinfo
  {pages} {4010} (\bibinfo {year} {2012})}\BibitemShut {NoStop}%
\bibitem [{\citenamefont {Wiklund}\ \emph {et~al.}(2012)\citenamefont
  {Wiklund}, \citenamefont {Green},\ and\ \citenamefont {Ohlin}}]{af_wiklund}%
  \BibitemOpen
  \bibfield  {author} {\bibinfo {author} {\bibfnamefont {M.}~\bibnamefont
  {Wiklund}}, \bibinfo {author} {\bibfnamefont {R.}~\bibnamefont {Green}},\
  and\ \bibinfo {author} {\bibfnamefont {M.}~\bibnamefont {Ohlin}},\ }\bibfield
   {title} {\bibinfo {title} {Acoustofluidics 14: Applications of acoustic
  streaming in microfluidic devices},\ }\href@noop {} {\bibfield  {journal}
  {\bibinfo  {journal} {Lab on a Chip}\ }\textbf {\bibinfo {volume} {12}},\
  \bibinfo {pages} {2438} (\bibinfo {year} {2012})}\BibitemShut {NoStop}%
\bibitem [{\citenamefont {Hartono}\ \emph {et~al.}(2011)\citenamefont
  {Hartono}, \citenamefont {Liu}, \citenamefont {Tan}, \citenamefont {Then},
  \citenamefont {Yung},\ and\ \citenamefont {Lim}}]{Lim_2011}%
  \BibitemOpen
  \bibfield  {author} {\bibinfo {author} {\bibfnamefont {D.}~\bibnamefont
  {Hartono}}, \bibinfo {author} {\bibfnamefont {Y.}~\bibnamefont {Liu}},
  \bibinfo {author} {\bibfnamefont {P.~L.}\ \bibnamefont {Tan}}, \bibinfo
  {author} {\bibfnamefont {X.~Y.~S.}\ \bibnamefont {Then}}, \bibinfo {author}
  {\bibfnamefont {L.-Y.~L.}\ \bibnamefont {Yung}},\ and\ \bibinfo {author}
  {\bibfnamefont {K.-M.}\ \bibnamefont {Lim}},\ }\bibfield  {title} {\bibinfo
  {title} {On-chip measurements of cell compressibility via acoustic
  radiation},\ }\href@noop {} {\bibfield  {journal} {\bibinfo  {journal} {Lab
  on a Chip}\ }\textbf {\bibinfo {volume} {11}},\ \bibinfo {pages} {4072}
  (\bibinfo {year} {2011})}\BibitemShut {NoStop}%
\bibitem [{\citenamefont {Mohapatra}\ \emph {et~al.}(2018)\citenamefont
  {Mohapatra}, \citenamefont {Sepehrirahnama},\ and\ \citenamefont
  {Lim}}]{Lim_2018}%
  \BibitemOpen
  \bibfield  {author} {\bibinfo {author} {\bibfnamefont {A.~R.}\ \bibnamefont
  {Mohapatra}}, \bibinfo {author} {\bibfnamefont {S.}~\bibnamefont
  {Sepehrirahnama}},\ and\ \bibinfo {author} {\bibfnamefont {K.-M.}\
  \bibnamefont {Lim}},\ }\bibfield  {title} {\bibinfo {title} {Experimental
  measurement of interparticle acoustic radiation force in the rayleigh
  limit},\ }\href@noop {} {\bibfield  {journal} {\bibinfo  {journal} {Physical
  Review E}\ }\textbf {\bibinfo {volume} {97}},\ \bibinfo {pages} {053105}
  (\bibinfo {year} {2018})}\BibitemShut {NoStop}%
\bibitem [{\citenamefont {Augustsson}\ \emph {et~al.}(2012)\citenamefont
  {Augustsson}, \citenamefont {Magnusson}, \citenamefont {Nordin},
  \citenamefont {Lilja},\ and\ \citenamefont {Laurell}}]{Laurell2012}%
  \BibitemOpen
  \bibfield  {author} {\bibinfo {author} {\bibfnamefont {P.}~\bibnamefont
  {Augustsson}}, \bibinfo {author} {\bibfnamefont {C.}~\bibnamefont
  {Magnusson}}, \bibinfo {author} {\bibfnamefont {M.}~\bibnamefont {Nordin}},
  \bibinfo {author} {\bibfnamefont {H.}~\bibnamefont {Lilja}},\ and\ \bibinfo
  {author} {\bibfnamefont {T.}~\bibnamefont {Laurell}},\ }\bibfield  {title}
  {\bibinfo {title} {Microfluidic, label-free enrichment of prostate cancer
  cells in blood based on acoustophoresis},\ }\href@noop {} {\bibfield
  {journal} {\bibinfo  {journal} {Analytical chemistry}\ }\textbf {\bibinfo
  {volume} {84}},\ \bibinfo {pages} {7954} (\bibinfo {year}
  {2012})}\BibitemShut {NoStop}%
\bibitem [{\citenamefont {Garcia-Sabat{\'e}}\ \emph {et~al.}(2014)\citenamefont
  {Garcia-Sabat{\'e}}, \citenamefont {Castro}, \citenamefont {Hoyos},\ and\
  \citenamefont {Gonz{\'a}lez-Cinca}}]{garcia2014experimental}%
  \BibitemOpen
  \bibfield  {author} {\bibinfo {author} {\bibfnamefont {A.}~\bibnamefont
  {Garcia-Sabat{\'e}}}, \bibinfo {author} {\bibfnamefont {A.}~\bibnamefont
  {Castro}}, \bibinfo {author} {\bibfnamefont {M.}~\bibnamefont {Hoyos}},\ and\
  \bibinfo {author} {\bibfnamefont {R.}~\bibnamefont {Gonz{\'a}lez-Cinca}},\
  }\bibfield  {title} {\bibinfo {title} {Experimental study on inter-particle
  acoustic forces},\ }\href@noop {} {\bibfield  {journal} {\bibinfo  {journal}
  {The Journal of the Acoustical Society of America}\ }\textbf {\bibinfo
  {volume} {135}},\ \bibinfo {pages} {1056} (\bibinfo {year}
  {2014})}\BibitemShut {NoStop}%
\bibitem [{\citenamefont {Bernassau}\ \emph {et~al.}(2014)\citenamefont
  {Bernassau}, \citenamefont {Glynne-Jones}, \citenamefont {Gesellchen},
  \citenamefont {Riehle}, \citenamefont {Hill},\ and\ \citenamefont
  {Cumming}}]{Hill_2014}%
  \BibitemOpen
  \bibfield  {author} {\bibinfo {author} {\bibfnamefont {A.}~\bibnamefont
  {Bernassau}}, \bibinfo {author} {\bibfnamefont {P.}~\bibnamefont
  {Glynne-Jones}}, \bibinfo {author} {\bibfnamefont {F.}~\bibnamefont
  {Gesellchen}}, \bibinfo {author} {\bibfnamefont {M.}~\bibnamefont {Riehle}},
  \bibinfo {author} {\bibfnamefont {M.}~\bibnamefont {Hill}},\ and\ \bibinfo
  {author} {\bibfnamefont {D.}~\bibnamefont {Cumming}},\ }\bibfield  {title}
  {\bibinfo {title} {Controlling acoustic streaming in an ultrasonic heptagonal
  tweezers with application to cell manipulation},\ }\href@noop {} {\bibfield
  {journal} {\bibinfo  {journal} {Ultrasonics}\ }\textbf {\bibinfo {volume}
  {54}},\ \bibinfo {pages} {268} (\bibinfo {year} {2014})}\BibitemShut
  {NoStop}%
\bibitem [{\citenamefont {Antfolk}\ \emph {et~al.}(2015)\citenamefont
  {Antfolk}, \citenamefont {Magnusson}, \citenamefont {Augustsson},
  \citenamefont {Lilja},\ and\ \citenamefont {Laurell}}]{antfolk2015_CTC}%
  \BibitemOpen
  \bibfield  {author} {\bibinfo {author} {\bibfnamefont {M.}~\bibnamefont
  {Antfolk}}, \bibinfo {author} {\bibfnamefont {C.}~\bibnamefont {Magnusson}},
  \bibinfo {author} {\bibfnamefont {P.}~\bibnamefont {Augustsson}}, \bibinfo
  {author} {\bibfnamefont {H.}~\bibnamefont {Lilja}},\ and\ \bibinfo {author}
  {\bibfnamefont {T.}~\bibnamefont {Laurell}},\ }\bibfield  {title} {\bibinfo
  {title} {Acoustofluidic, label-free separation and simultaneous concentration
  of rare tumor cells from white blood cells},\ }\href@noop {} {\bibfield
  {journal} {\bibinfo  {journal} {Analytical chemistry}\ }\textbf {\bibinfo
  {volume} {87}},\ \bibinfo {pages} {9322} (\bibinfo {year}
  {2015})}\BibitemShut {NoStop}%
\bibitem [{\citenamefont {Marzo}\ \emph {et~al.}(2015)\citenamefont {Marzo},
  \citenamefont {Seah}, \citenamefont {Drinkwater}, \citenamefont {Sahoo},
  \citenamefont {Long},\ and\ \citenamefont {Subramanian}}]{DW2015}%
  \BibitemOpen
  \bibfield  {author} {\bibinfo {author} {\bibfnamefont {A.}~\bibnamefont
  {Marzo}}, \bibinfo {author} {\bibfnamefont {S.~A.}\ \bibnamefont {Seah}},
  \bibinfo {author} {\bibfnamefont {B.~W.}\ \bibnamefont {Drinkwater}},
  \bibinfo {author} {\bibfnamefont {D.~R.}\ \bibnamefont {Sahoo}}, \bibinfo
  {author} {\bibfnamefont {B.}~\bibnamefont {Long}},\ and\ \bibinfo {author}
  {\bibfnamefont {S.}~\bibnamefont {Subramanian}},\ }\bibfield  {title}
  {\bibinfo {title} {Holographic acoustic elements for manipulation of
  levitated objects},\ }\href@noop {} {\bibfield  {journal} {\bibinfo
  {journal} {Nature communications}\ }\textbf {\bibinfo {volume} {6}},\
  \bibinfo {pages} {8661} (\bibinfo {year} {2015})}\BibitemShut {NoStop}%
\bibitem [{\citenamefont {Wijaya}\ \emph {et~al.}(2016)\citenamefont {Wijaya},
  \citenamefont {Mohapatra}, \citenamefont {Sepehrirahnama},\ and\
  \citenamefont {Lim}}]{Lim_2016}%
  \BibitemOpen
  \bibfield  {author} {\bibinfo {author} {\bibfnamefont {F.~B.}\ \bibnamefont
  {Wijaya}}, \bibinfo {author} {\bibfnamefont {A.~R.}\ \bibnamefont
  {Mohapatra}}, \bibinfo {author} {\bibfnamefont {S.}~\bibnamefont
  {Sepehrirahnama}},\ and\ \bibinfo {author} {\bibfnamefont {K.-M.}\
  \bibnamefont {Lim}},\ }\bibfield  {title} {\bibinfo {title} {Coupled
  acoustic-shell model for experimental study of cell stiffness under
  acoustophoresis},\ }\href@noop {} {\bibfield  {journal} {\bibinfo  {journal}
  {Microfluidics and Nanofluidics}\ }\textbf {\bibinfo {volume} {20}},\
  \bibinfo {pages} {69} (\bibinfo {year} {2016})}\BibitemShut {NoStop}%
\bibitem [{\citenamefont {Marzo}\ and\ \citenamefont
  {Drinkwater}(2019)}]{DW2019}%
  \BibitemOpen
  \bibfield  {author} {\bibinfo {author} {\bibfnamefont {A.}~\bibnamefont
  {Marzo}}\ and\ \bibinfo {author} {\bibfnamefont {B.~W.}\ \bibnamefont
  {Drinkwater}},\ }\bibfield  {title} {\bibinfo {title} {Holographic acoustic
  tweezers},\ }\href@noop {} {\bibfield  {journal} {\bibinfo  {journal}
  {Proceedings of the National Academy of Sciences}\ }\textbf {\bibinfo
  {volume} {116}},\ \bibinfo {pages} {84} (\bibinfo {year} {2019})}\BibitemShut
  {NoStop}%
\bibitem [{\citenamefont {Polychronopoulos}\ and\ \citenamefont
  {Memoli}(2020)}]{memoli2020AcLevMeta}%
  \BibitemOpen
  \bibfield  {author} {\bibinfo {author} {\bibfnamefont {S.}~\bibnamefont
  {Polychronopoulos}}\ and\ \bibinfo {author} {\bibfnamefont {G.}~\bibnamefont
  {Memoli}},\ }\bibfield  {title} {\bibinfo {title} {Acoustic levitation with
  optimized reflective metamaterials},\ }\href
  {https://doi.org/10.1038/s41598-020-60978-4} {\bibfield  {journal} {\bibinfo
  {journal} {Scientific reports}\ }\textbf {\bibinfo {volume} {10}},\ \bibinfo
  {pages} {1} (\bibinfo {year} {2020})}\BibitemShut {NoStop}%
\bibitem [{\citenamefont {King}(1934)}]{King_34}%
  \BibitemOpen
  \bibfield  {author} {\bibinfo {author} {\bibfnamefont {L.~V.}\ \bibnamefont
  {King}},\ }\bibfield  {title} {\bibinfo {title} {On the acoustic radiation
  pressure on spheres},\ }\href@noop {} {\bibfield  {journal} {\bibinfo
  {journal} {Proceedings of the Royal Society of London. Series A-Mathematical
  and Physical Sciences}\ }\textbf {\bibinfo {volume} {147}},\ \bibinfo {pages}
  {212} (\bibinfo {year} {1934})}\BibitemShut {NoStop}%
\bibitem [{\citenamefont {Yosioka}\ and\ \citenamefont
  {Kawasima}(1955)}]{Yosioka_55}%
  \BibitemOpen
  \bibfield  {author} {\bibinfo {author} {\bibfnamefont {K.}~\bibnamefont
  {Yosioka}}\ and\ \bibinfo {author} {\bibfnamefont {Y.}~\bibnamefont
  {Kawasima}},\ }\bibfield  {title} {\bibinfo {title} {Acoustic radiation
  pressure on a compressible sphere},\ }\href@noop {} {\bibfield  {journal}
  {\bibinfo  {journal} {Acta Acustica united with Acustica}\ }\textbf {\bibinfo
  {volume} {5}},\ \bibinfo {pages} {167} (\bibinfo {year} {1955})}\BibitemShut
  {NoStop}%
\bibitem [{\citenamefont {Gorkov}(1962)}]{Gorkov_62}%
  \BibitemOpen
  \bibfield  {author} {\bibinfo {author} {\bibfnamefont {L.~P.}\ \bibnamefont
  {Gorkov}},\ }\bibfield  {title} {\bibinfo {title} {On the forces acting on a
  small particle in an acoustic filed in an ideal fluid},\ }\href@noop {}
  {\bibfield  {journal} {\bibinfo  {journal} {Soviet Physics - Doklady}\
  }\textbf {\bibinfo {volume} {6}},\ \bibinfo {pages} {773} (\bibinfo {year}
  {1962})}\BibitemShut {NoStop}%
\bibitem [{\citenamefont {Sepehrirahnama}\ \emph
  {et~al.}(2015{\natexlab{a}})\citenamefont {Sepehrirahnama}, \citenamefont
  {Lim},\ and\ \citenamefont {Chau}}]{Shahrokh_2015}%
  \BibitemOpen
  \bibfield  {author} {\bibinfo {author} {\bibfnamefont {S.}~\bibnamefont
  {Sepehrirahnama}}, \bibinfo {author} {\bibfnamefont {K.-M.}\ \bibnamefont
  {Lim}},\ and\ \bibinfo {author} {\bibfnamefont {F.~S.}\ \bibnamefont
  {Chau}},\ }\bibfield  {title} {\bibinfo {title} {Numerical analysis of the
  acoustic radiation force and acoustic streaming around a sphere in an
  acoustic standing wave},\ }\href@noop {} {\bibfield  {journal} {\bibinfo
  {journal} {Physics Procedia}\ }\textbf {\bibinfo {volume} {70}},\ \bibinfo
  {pages} {80} (\bibinfo {year} {2015}{\natexlab{a}})}\BibitemShut {NoStop}%
\bibitem [{\citenamefont {Silva}\ and\ \citenamefont
  {Bruus}(2014)}]{silva_Bruus}%
  \BibitemOpen
  \bibfield  {author} {\bibinfo {author} {\bibfnamefont {G.~T.}\ \bibnamefont
  {Silva}}\ and\ \bibinfo {author} {\bibfnamefont {H.}~\bibnamefont {Bruus}},\
  }\bibfield  {title} {\bibinfo {title} {Acoustic interaction forces between
  small particles in an ideal fluid},\ }\href@noop {} {\bibfield  {journal}
  {\bibinfo  {journal} {Physical Review E}\ }\textbf {\bibinfo {volume} {90}},\
  \bibinfo {pages} {063007} (\bibinfo {year} {2014})}\BibitemShut {NoStop}%
\bibitem [{\citenamefont {Lopes}\ \emph {et~al.}(2016)\citenamefont {Lopes},
  \citenamefont {Azarpeyvand},\ and\ \citenamefont
  {Silva}}]{lopes2016acoustic}%
  \BibitemOpen
  \bibfield  {author} {\bibinfo {author} {\bibfnamefont {J.~H.}\ \bibnamefont
  {Lopes}}, \bibinfo {author} {\bibfnamefont {M.}~\bibnamefont {Azarpeyvand}},\
  and\ \bibinfo {author} {\bibfnamefont {G.~T.}\ \bibnamefont {Silva}},\
  }\bibfield  {title} {\bibinfo {title} {Acoustic interaction forces and
  torques acting on suspended spheres in an ideal fluid},\ }\href@noop {}
  {\bibfield  {journal} {\bibinfo  {journal} {IEEE transactions on ultrasonics,
  ferroelectrics, and frequency control}\ }\textbf {\bibinfo {volume} {63}},\
  \bibinfo {pages} {186} (\bibinfo {year} {2016})}\BibitemShut {NoStop}%
\bibitem [{\citenamefont {Sepehrirahnama}\ \emph {et~al.}(2016)\citenamefont
  {Sepehrirahnama}, \citenamefont {Chau},\ and\ \citenamefont
  {Lim}}]{Shahrokh_2016}%
  \BibitemOpen
  \bibfield  {author} {\bibinfo {author} {\bibfnamefont {S.}~\bibnamefont
  {Sepehrirahnama}}, \bibinfo {author} {\bibfnamefont {F.~S.}\ \bibnamefont
  {Chau}},\ and\ \bibinfo {author} {\bibfnamefont {K.-M.}\ \bibnamefont
  {Lim}},\ }\bibfield  {title} {\bibinfo {title} {Effects of viscosity and
  acoustic streaming on the interparticle radiation force between rigid spheres
  in a standing wave},\ }\href@noop {} {\bibfield  {journal} {\bibinfo
  {journal} {Physical Review E}\ }\textbf {\bibinfo {volume} {93}},\ \bibinfo
  {pages} {023307} (\bibinfo {year} {2016})}\BibitemShut {NoStop}%
\bibitem [{\citenamefont {Sepehrirahnama}\ and\ \citenamefont
  {Lim}(2020{\natexlab{a}})}]{shahrokh2020PRE}%
  \BibitemOpen
  \bibfield  {author} {\bibinfo {author} {\bibfnamefont {S.}~\bibnamefont
  {Sepehrirahnama}}\ and\ \bibinfo {author} {\bibfnamefont {K.-M.}\
  \bibnamefont {Lim}},\ }\bibfield  {title} {\bibinfo {title} {Generalized
  potential theory for close-range acoustic interactions in the rayleigh
  limit},\ }\href {https://doi.org/10.1103/PhysRevE.102.043307} {\bibfield
  {journal} {\bibinfo  {journal} {Physical Review E}\ }\textbf {\bibinfo
  {volume} {102}},\ \bibinfo {pages} {043307} (\bibinfo {year}
  {2020}{\natexlab{a}})}\BibitemShut {NoStop}%
\bibitem [{\citenamefont {Sepehrirahnama}\ and\ \citenamefont
  {Lim}(2020{\natexlab{b}})}]{Sh2020_agglo}%
  \BibitemOpen
  \bibfield  {author} {\bibinfo {author} {\bibfnamefont {S.}~\bibnamefont
  {Sepehrirahnama}}\ and\ \bibinfo {author} {\bibfnamefont {K.-M.}\
  \bibnamefont {Lim}},\ }\bibfield  {title} {\bibinfo {title} {Acoustophoretic
  agglomeration patterns of particulate phase in a host fluid},\ }\href@noop {}
  {\bibfield  {journal} {\bibinfo  {journal} {Microfluidics and Nanofluidics}\
  }\textbf {\bibinfo {volume} {24}},\ \bibinfo {pages} {1} (\bibinfo {year}
  {2020}{\natexlab{b}})}\BibitemShut {NoStop}%
\bibitem [{\citenamefont {Doinikov}(2001)}]{Doinikov2001_interparticle}%
  \BibitemOpen
  \bibfield  {author} {\bibinfo {author} {\bibfnamefont {A.~A.}\ \bibnamefont
  {Doinikov}},\ }\bibfield  {title} {\bibinfo {title} {Acoustic radiation
  interparticle forces in a compressible fluid},\ }\href@noop {} {\bibfield
  {journal} {\bibinfo  {journal} {Journal of Fluid Mechanics}\ }\textbf
  {\bibinfo {volume} {444}},\ \bibinfo {pages} {1} (\bibinfo {year}
  {2001})}\BibitemShut {NoStop}%
\bibitem [{\citenamefont {Fan}\ \emph {et~al.}(2008)\citenamefont {Fan},
  \citenamefont {Mei}, \citenamefont {Yang},\ and\ \citenamefont
  {Chen}}]{fan2008torque}%
  \BibitemOpen
  \bibfield  {author} {\bibinfo {author} {\bibfnamefont {Z.}~\bibnamefont
  {Fan}}, \bibinfo {author} {\bibfnamefont {D.}~\bibnamefont {Mei}}, \bibinfo
  {author} {\bibfnamefont {K.}~\bibnamefont {Yang}},\ and\ \bibinfo {author}
  {\bibfnamefont {Z.}~\bibnamefont {Chen}},\ }\bibfield  {title} {\bibinfo
  {title} {Acoustic radiation torque on an irregularly shaped scatterer in an
  arbitrary sound field},\ }\href@noop {} {\bibfield  {journal} {\bibinfo
  {journal} {The Journal of the Acoustical Society of America}\ }\textbf
  {\bibinfo {volume} {124}},\ \bibinfo {pages} {2727} (\bibinfo {year}
  {2008})}\BibitemShut {NoStop}%
\bibitem [{\citenamefont {Zhang}\ and\ \citenamefont
  {Marston}(2011)}]{zhang2011torque}%
  \BibitemOpen
  \bibfield  {author} {\bibinfo {author} {\bibfnamefont {L.}~\bibnamefont
  {Zhang}}\ and\ \bibinfo {author} {\bibfnamefont {P.~L.}\ \bibnamefont
  {Marston}},\ }\bibfield  {title} {\bibinfo {title} {Acoustic radiation torque
  and the conservation of angular momentum (l)},\ }\href@noop {} {\bibfield
  {journal} {\bibinfo  {journal} {The Journal of the Acoustical Society of
  America}\ }\textbf {\bibinfo {volume} {129}},\ \bibinfo {pages} {1679}
  (\bibinfo {year} {2011})}\BibitemShut {NoStop}%
\bibitem [{\citenamefont {Sepehrirahnama}\ \emph
  {et~al.}(2015{\natexlab{b}})\citenamefont {Sepehrirahnama}, \citenamefont
  {Lim},\ and\ \citenamefont {Chau}}]{Shahrokh_2015_main}%
  \BibitemOpen
  \bibfield  {author} {\bibinfo {author} {\bibfnamefont {S.}~\bibnamefont
  {Sepehrirahnama}}, \bibinfo {author} {\bibfnamefont {K.-M.}\ \bibnamefont
  {Lim}},\ and\ \bibinfo {author} {\bibfnamefont {F.~S.}\ \bibnamefont
  {Chau}},\ }\bibfield  {title} {\bibinfo {title} {Numerical study of
  interparticle radiation force acting on rigid spheres in a standing wave},\
  }\href@noop {} {\bibfield  {journal} {\bibinfo  {journal} {The Journal of the
  Acoustical Society of America}\ }\textbf {\bibinfo {volume} {137}},\ \bibinfo
  {pages} {2614} (\bibinfo {year} {2015}{\natexlab{b}})}\BibitemShut {NoStop}%
\bibitem [{\citenamefont {Doinikov}(1994{\natexlab{a}})}]{Doinikov1994_JFM}%
  \BibitemOpen
  \bibfield  {author} {\bibinfo {author} {\bibfnamefont {A.~A.}\ \bibnamefont
  {Doinikov}},\ }\bibfield  {title} {\bibinfo {title} {Acoustic radiation
  pressure on a compressible sphere in a viscous fluid},\ }\href@noop {}
  {\bibfield  {journal} {\bibinfo  {journal} {Journal of Fluid Mechanics}\
  }\textbf {\bibinfo {volume} {267}},\ \bibinfo {pages} {1} (\bibinfo {year}
  {1994}{\natexlab{a}})}\BibitemShut {NoStop}%
\bibitem [{\citenamefont {Doinikov}(1994{\natexlab{b}})}]{Doinikov1994_Proc}%
  \BibitemOpen
  \bibfield  {author} {\bibinfo {author} {\bibfnamefont {A.~A.}\ \bibnamefont
  {Doinikov}},\ }\bibfield  {title} {\bibinfo {title} {Acoustic radiation
  pressure on a rigid sphere in a viscous fluid},\ }\href@noop {} {\bibfield
  {journal} {\bibinfo  {journal} {Proceedings of the Royal Society of London.
  Series A: Mathematical and Physical Sciences}\ }\textbf {\bibinfo {volume}
  {447}},\ \bibinfo {pages} {447} (\bibinfo {year}
  {1994}{\natexlab{b}})}\BibitemShut {NoStop}%
\bibitem [{\citenamefont {Foresti}\ \emph {et~al.}(2012)\citenamefont
  {Foresti}, \citenamefont {Nabavi},\ and\ \citenamefont
  {Poulikakos}}]{foresti2012ellSpheroid}%
  \BibitemOpen
  \bibfield  {author} {\bibinfo {author} {\bibfnamefont {D.}~\bibnamefont
  {Foresti}}, \bibinfo {author} {\bibfnamefont {M.}~\bibnamefont {Nabavi}},\
  and\ \bibinfo {author} {\bibfnamefont {D.}~\bibnamefont {Poulikakos}},\
  }\bibfield  {title} {\bibinfo {title} {On the acoustic levitation stability
  behaviour of spherical and ellipsoidal particles},\ }\href@noop {} {\bibfield
   {journal} {\bibinfo  {journal} {Journal of Fluid Mechanics}\ }\textbf
  {\bibinfo {volume} {709}},\ \bibinfo {pages} {581} (\bibinfo {year}
  {2012})}\BibitemShut {NoStop}%
\bibitem [{\citenamefont {Wijaya}\ and\ \citenamefont
  {Lim}(2015)}]{FBW2015spheroid}%
  \BibitemOpen
  \bibfield  {author} {\bibinfo {author} {\bibfnamefont {F.~B.}\ \bibnamefont
  {Wijaya}}\ and\ \bibinfo {author} {\bibfnamefont {K.-M.}\ \bibnamefont
  {Lim}},\ }\bibfield  {title} {\bibinfo {title} {Numerical calculation of
  acoustic radiation force and torque acting on rigid non-spherical
  particles},\ }\href@noop {} {\bibfield  {journal} {\bibinfo  {journal} {Acta
  Acustica united with Acustica}\ }\textbf {\bibinfo {volume} {101}},\ \bibinfo
  {pages} {531} (\bibinfo {year} {2015})}\BibitemShut {NoStop}%
\bibitem [{\citenamefont {Mitri}(2015{\natexlab{a}})}]{mitri2015ellCyl}%
  \BibitemOpen
  \bibfield  {author} {\bibinfo {author} {\bibfnamefont {F.}~\bibnamefont
  {Mitri}},\ }\bibfield  {title} {\bibinfo {title} {Acoustic radiation force on
  a rigid elliptical cylinder in plane (quasi) standing waves},\ }\href@noop {}
  {\bibfield  {journal} {\bibinfo  {journal} {Journal of Applied Physics}\
  }\textbf {\bibinfo {volume} {118}},\ \bibinfo {pages} {214903} (\bibinfo
  {year} {2015}{\natexlab{a}})}\BibitemShut {NoStop}%
\bibitem [{\citenamefont {Wei}\ \emph {et~al.}(2004)\citenamefont {Wei},
  \citenamefont {Thiessen},\ and\ \citenamefont {Marston}}]{wei2004cyl}%
  \BibitemOpen
  \bibfield  {author} {\bibinfo {author} {\bibfnamefont {W.}~\bibnamefont
  {Wei}}, \bibinfo {author} {\bibfnamefont {D.~B.}\ \bibnamefont {Thiessen}},\
  and\ \bibinfo {author} {\bibfnamefont {P.~L.}\ \bibnamefont {Marston}},\
  }\bibfield  {title} {\bibinfo {title} {Acoustic radiation force on a
  compressible cylinder in a standing wave},\ }\href@noop {} {\bibfield
  {journal} {\bibinfo  {journal} {The Journal of the Acoustical Society of
  America}\ }\textbf {\bibinfo {volume} {116}},\ \bibinfo {pages} {201}
  (\bibinfo {year} {2004})}\BibitemShut {NoStop}%
\bibitem [{\citenamefont {Xie}\ and\ \citenamefont
  {Wei}(2004)}]{xie2004ARFdisc}%
  \BibitemOpen
  \bibfield  {author} {\bibinfo {author} {\bibfnamefont {W.}~\bibnamefont
  {Xie}}\ and\ \bibinfo {author} {\bibfnamefont {B.}~\bibnamefont {Wei}},\
  }\bibfield  {title} {\bibinfo {title} {Dynamics of acoustically levitated
  disk samples},\ }\href@noop {} {\bibfield  {journal} {\bibinfo  {journal}
  {Physical Review E}\ }\textbf {\bibinfo {volume} {70}},\ \bibinfo {pages}
  {046611} (\bibinfo {year} {2004})}\BibitemShut {NoStop}%
\bibitem [{\citenamefont {Garbin}\ \emph {et~al.}(2015)\citenamefont {Garbin},
  \citenamefont {Leibacher}, \citenamefont {Hahn}, \citenamefont {Le~Ferrand},
  \citenamefont {Studart},\ and\ \citenamefont {Dual}}]{garbin2015ARFdisk}%
  \BibitemOpen
  \bibfield  {author} {\bibinfo {author} {\bibfnamefont {A.}~\bibnamefont
  {Garbin}}, \bibinfo {author} {\bibfnamefont {I.}~\bibnamefont {Leibacher}},
  \bibinfo {author} {\bibfnamefont {P.}~\bibnamefont {Hahn}}, \bibinfo {author}
  {\bibfnamefont {H.}~\bibnamefont {Le~Ferrand}}, \bibinfo {author}
  {\bibfnamefont {A.}~\bibnamefont {Studart}},\ and\ \bibinfo {author}
  {\bibfnamefont {J.}~\bibnamefont {Dual}},\ }\bibfield  {title} {\bibinfo
  {title} {Acoustophoresis of disk-shaped microparticles: A numerical and
  experimental study of acoustic radiation forces and torques},\ }\href@noop {}
  {\bibfield  {journal} {\bibinfo  {journal} {The Journal of the Acoustical
  Society of America}\ }\textbf {\bibinfo {volume} {138}},\ \bibinfo {pages}
  {2759} (\bibinfo {year} {2015})}\BibitemShut {NoStop}%
\bibitem [{\citenamefont {Wijaya}\ \emph {et~al.}(2018)\citenamefont {Wijaya},
  \citenamefont {Sepehrirahnama},\ and\ \citenamefont {Lim}}]{Lim_2018b}%
  \BibitemOpen
  \bibfield  {author} {\bibinfo {author} {\bibfnamefont {F.~B.}\ \bibnamefont
  {Wijaya}}, \bibinfo {author} {\bibfnamefont {S.}~\bibnamefont
  {Sepehrirahnama}},\ and\ \bibinfo {author} {\bibfnamefont {K.-M.}\
  \bibnamefont {Lim}},\ }\bibfield  {title} {\bibinfo {title} {Interparticle
  force and torque on rigid spheroidal particles in acoustophoresis},\
  }\href@noop {} {\bibfield  {journal} {\bibinfo  {journal} {Wave Motion}\
  }\textbf {\bibinfo {volume} {81}},\ \bibinfo {pages} {28} (\bibinfo {year}
  {2018})}\BibitemShut {NoStop}%
\bibitem [{\citenamefont {Sieck}\ \emph {et~al.}(2017)\citenamefont {Sieck},
  \citenamefont {Al{\`u}},\ and\ \citenamefont {Haberman}}]{Alu2017WCorigin}%
  \BibitemOpen
  \bibfield  {author} {\bibinfo {author} {\bibfnamefont {C.~F.}\ \bibnamefont
  {Sieck}}, \bibinfo {author} {\bibfnamefont {A.}~\bibnamefont {Al{\`u}}},\
  and\ \bibinfo {author} {\bibfnamefont {M.~R.}\ \bibnamefont {Haberman}},\
  }\bibfield  {title} {\bibinfo {title} {Origins of willis coupling and
  acoustic bianisotropy in acoustic metamaterials through source-driven
  homogenization},\ }\href@noop {} {\bibfield  {journal} {\bibinfo  {journal}
  {Physical Review B}\ }\textbf {\bibinfo {volume} {96}},\ \bibinfo {pages}
  {104303} (\bibinfo {year} {2017})}\BibitemShut {NoStop}%
\bibitem [{\citenamefont {Quan}\ \emph {et~al.}(2018)\citenamefont {Quan},
  \citenamefont {Ra’di}, \citenamefont {Sounas},\ and\ \citenamefont
  {Al{\`u}}}]{Alu2018maxWC}%
  \BibitemOpen
  \bibfield  {author} {\bibinfo {author} {\bibfnamefont {L.}~\bibnamefont
  {Quan}}, \bibinfo {author} {\bibfnamefont {Y.}~\bibnamefont {Ra’di}},
  \bibinfo {author} {\bibfnamefont {D.~L.}\ \bibnamefont {Sounas}},\ and\
  \bibinfo {author} {\bibfnamefont {A.}~\bibnamefont {Al{\`u}}},\ }\bibfield
  {title} {\bibinfo {title} {Maximum willis coupling in acoustic scatterers},\
  }\href@noop {} {\bibfield  {journal} {\bibinfo  {journal} {Physical Review
  Letters}\ }\textbf {\bibinfo {volume} {120}},\ \bibinfo {pages} {254301}
  (\bibinfo {year} {2018})}\BibitemShut {NoStop}%
\bibitem [{\citenamefont {Jordaan}\ \emph {et~al.}(2018)\citenamefont
  {Jordaan}, \citenamefont {Punzet}, \citenamefont {Melnikov}, \citenamefont
  {Sanches}, \citenamefont {Oberst}, \citenamefont {Marburg},\ and\
  \citenamefont {Powell}}]{jordaan2018}%
  \BibitemOpen
  \bibfield  {author} {\bibinfo {author} {\bibfnamefont {J.}~\bibnamefont
  {Jordaan}}, \bibinfo {author} {\bibfnamefont {S.}~\bibnamefont {Punzet}},
  \bibinfo {author} {\bibfnamefont {A.}~\bibnamefont {Melnikov}}, \bibinfo
  {author} {\bibfnamefont {A.}~\bibnamefont {Sanches}}, \bibinfo {author}
  {\bibfnamefont {S.}~\bibnamefont {Oberst}}, \bibinfo {author} {\bibfnamefont
  {S.}~\bibnamefont {Marburg}},\ and\ \bibinfo {author} {\bibfnamefont {D.~A.}\
  \bibnamefont {Powell}},\ }\bibfield  {title} {\bibinfo {title} {Measuring
  monopole and dipole polarizability of acoustic meta-atoms},\ }\href@noop {}
  {\bibfield  {journal} {\bibinfo  {journal} {Applied Physics Letters}\
  }\textbf {\bibinfo {volume} {113}},\ \bibinfo {pages} {224102} (\bibinfo
  {year} {2018})}\BibitemShut {NoStop}%
\bibitem [{\citenamefont {Melnikov}\ \emph {et~al.}(2019)\citenamefont
  {Melnikov}, \citenamefont {Chiang}, \citenamefont {Quan}, \citenamefont
  {Oberst}, \citenamefont {Al{\`u}}, \citenamefont {Marburg},\ and\
  \citenamefont {Powell}}]{anton2019}%
  \BibitemOpen
  \bibfield  {author} {\bibinfo {author} {\bibfnamefont {A.}~\bibnamefont
  {Melnikov}}, \bibinfo {author} {\bibfnamefont {Y.~K.}\ \bibnamefont
  {Chiang}}, \bibinfo {author} {\bibfnamefont {L.}~\bibnamefont {Quan}},
  \bibinfo {author} {\bibfnamefont {S.}~\bibnamefont {Oberst}}, \bibinfo
  {author} {\bibfnamefont {A.}~\bibnamefont {Al{\`u}}}, \bibinfo {author}
  {\bibfnamefont {S.}~\bibnamefont {Marburg}},\ and\ \bibinfo {author}
  {\bibfnamefont {D.}~\bibnamefont {Powell}},\ }\bibfield  {title} {\bibinfo
  {title} {Acoustic meta-atom with experimentally verified maximum willis
  coupling},\ }\href@noop {} {\bibfield  {journal} {\bibinfo  {journal} {Nature
  communications}\ }\textbf {\bibinfo {volume} {10}},\ \bibinfo {pages} {1}
  (\bibinfo {year} {2019})}\BibitemShut {NoStop}%
\bibitem [{\citenamefont {Chiang}\ \emph {et~al.}(2020)\citenamefont {Chiang},
  \citenamefont {Oberst}, \citenamefont {Melnikov}, \citenamefont {Quan},
  \citenamefont {Marburg}, \citenamefont {Al{\`u}},\ and\ \citenamefont
  {Powell}}]{YK2020arraye}%
  \BibitemOpen
  \bibfield  {author} {\bibinfo {author} {\bibfnamefont {Y.~K.}\ \bibnamefont
  {Chiang}}, \bibinfo {author} {\bibfnamefont {S.}~\bibnamefont {Oberst}},
  \bibinfo {author} {\bibfnamefont {A.}~\bibnamefont {Melnikov}}, \bibinfo
  {author} {\bibfnamefont {L.}~\bibnamefont {Quan}}, \bibinfo {author}
  {\bibfnamefont {S.}~\bibnamefont {Marburg}}, \bibinfo {author} {\bibfnamefont
  {A.}~\bibnamefont {Al{\`u}}},\ and\ \bibinfo {author} {\bibfnamefont {D.~A.}\
  \bibnamefont {Powell}},\ }\bibfield  {title} {\bibinfo {title}
  {Reconfigurable acoustic metagrating for high-efficiency anomalous
  reflection},\ }\href@noop {} {\bibfield  {journal} {\bibinfo  {journal}
  {Physical Review Applied}\ }\textbf {\bibinfo {volume} {13}},\ \bibinfo
  {pages} {064067} (\bibinfo {year} {2020})}\BibitemShut {NoStop}%
\bibitem [{\citenamefont {Su}\ and\ \citenamefont {Norris}(2018)}]{norris2018}%
  \BibitemOpen
  \bibfield  {author} {\bibinfo {author} {\bibfnamefont {X.}~\bibnamefont
  {Su}}\ and\ \bibinfo {author} {\bibfnamefont {A.~N.}\ \bibnamefont
  {Norris}},\ }\bibfield  {title} {\bibinfo {title} {Retrieval method for the
  bianisotropic polarizability tensor of willis acoustic scatterers},\
  }\href@noop {} {\bibfield  {journal} {\bibinfo  {journal} {Physical Review
  B}\ }\textbf {\bibinfo {volume} {98}},\ \bibinfo {pages} {174305} (\bibinfo
  {year} {2018})}\BibitemShut {NoStop}%
\bibitem [{\citenamefont {Settnes}\ and\ \citenamefont
  {Bruus}(2012)}]{Bruus_2012}%
  \BibitemOpen
  \bibfield  {author} {\bibinfo {author} {\bibfnamefont {M.}~\bibnamefont
  {Settnes}}\ and\ \bibinfo {author} {\bibfnamefont {H.}~\bibnamefont
  {Bruus}},\ }\bibfield  {title} {\bibinfo {title} {Forces acting on a small
  particle in an acoustical field in a viscous fluid},\ }\href@noop {}
  {\bibfield  {journal} {\bibinfo  {journal} {Physical Review E}\ }\textbf
  {\bibinfo {volume} {85}},\ \bibinfo {pages} {016327} (\bibinfo {year}
  {2012})}\BibitemShut {NoStop}%
\bibitem [{\citenamefont {Maidanik}(1958)}]{maidanik1958torque}%
  \BibitemOpen
  \bibfield  {author} {\bibinfo {author} {\bibfnamefont {G.}~\bibnamefont
  {Maidanik}},\ }\bibfield  {title} {\bibinfo {title} {Torques due to
  acoustical radiation pressure},\ }\href@noop {} {\bibfield  {journal}
  {\bibinfo  {journal} {The Journal of the Acoustical Society of America}\
  }\textbf {\bibinfo {volume} {30}},\ \bibinfo {pages} {620} (\bibinfo {year}
  {1958})}\BibitemShut {NoStop}%
\bibitem [{\citenamefont {Toftul}\ \emph {et~al.}(2019)\citenamefont {Toftul},
  \citenamefont {Bliokh}, \citenamefont {Petrov},\ and\ \citenamefont
  {Nori}}]{toftul2019CanonicalForm}%
  \BibitemOpen
  \bibfield  {author} {\bibinfo {author} {\bibfnamefont {I.}~\bibnamefont
  {Toftul}}, \bibinfo {author} {\bibfnamefont {K.}~\bibnamefont {Bliokh}},
  \bibinfo {author} {\bibfnamefont {M.~I.}\ \bibnamefont {Petrov}},\ and\
  \bibinfo {author} {\bibfnamefont {F.}~\bibnamefont {Nori}},\ }\bibfield
  {title} {\bibinfo {title} {Acoustic radiation force and torque on small
  particles as measures of the canonical momentum and spin densities},\
  }\href@noop {} {\bibfield  {journal} {\bibinfo  {journal} {Physical review
  letters}\ }\textbf {\bibinfo {volume} {123}},\ \bibinfo {pages} {183901}
  (\bibinfo {year} {2019})}\BibitemShut {NoStop}%
\bibitem [{\citenamefont {Sepehrirahnama}\ \emph
  {et~al.}(2015{\natexlab{c}})\citenamefont {Sepehrirahnama}, \citenamefont
  {Chau},\ and\ \citenamefont {Lim}}]{sepehrirahnama2015numerical}%
  \BibitemOpen
  \bibfield  {author} {\bibinfo {author} {\bibfnamefont {S.}~\bibnamefont
  {Sepehrirahnama}}, \bibinfo {author} {\bibfnamefont {F.~S.}\ \bibnamefont
  {Chau}},\ and\ \bibinfo {author} {\bibfnamefont {K.-M.}\ \bibnamefont
  {Lim}},\ }\bibfield  {title} {\bibinfo {title} {Numerical calculation of
  acoustic radiation forces acting on a sphere in a viscous fluid},\
  }\href@noop {} {\bibfield  {journal} {\bibinfo  {journal} {Physical Review
  E}\ }\textbf {\bibinfo {volume} {92}},\ \bibinfo {pages} {063309} (\bibinfo
  {year} {2015}{\natexlab{c}})}\BibitemShut {NoStop}%
\bibitem [{\citenamefont {Hasegawa}\ and\ \citenamefont
  {Yosioka}(1969)}]{Hasegawa_69}%
  \BibitemOpen
  \bibfield  {author} {\bibinfo {author} {\bibfnamefont {T.}~\bibnamefont
  {Hasegawa}}\ and\ \bibinfo {author} {\bibfnamefont {K.}~\bibnamefont
  {Yosioka}},\ }\bibfield  {title} {\bibinfo {title} {Acoustic-radiation force
  on a solid elastic sphere},\ }\href@noop {} {\bibfield  {journal} {\bibinfo
  {journal} {The Journal of the Acoustical Society of America}\ }\textbf
  {\bibinfo {volume} {46}},\ \bibinfo {pages} {1139} (\bibinfo {year}
  {1969})}\BibitemShut {NoStop}%
\bibitem [{\citenamefont {Marston}\ \emph {et~al.}(2006)\citenamefont
  {Marston}, \citenamefont {Wei},\ and\ \citenamefont {Thiessen}}]{Marston}%
  \BibitemOpen
  \bibfield  {author} {\bibinfo {author} {\bibfnamefont {P.~L.}\ \bibnamefont
  {Marston}}, \bibinfo {author} {\bibfnamefont {W.}~\bibnamefont {Wei}},\ and\
  \bibinfo {author} {\bibfnamefont {D.~B.}\ \bibnamefont {Thiessen}},\
  }\bibfield  {title} {\bibinfo {title} {Acoustic radiation force on elliptical
  cylinders and spheroidal objects in low frequency standing waves},\ }in\
  \href@noop {} {\emph {\bibinfo {booktitle} {AIP Conference Proceedings}}},\
  Vol.\ \bibinfo {volume} {838}\ (\bibinfo {organization} {AIP},\ \bibinfo
  {year} {2006})\ pp.\ \bibinfo {pages} {495--499}\BibitemShut {NoStop}%
\bibitem [{\citenamefont {Silva}(2011)}]{silva2011_bessel_beam_offaxis}%
  \BibitemOpen
  \bibfield  {author} {\bibinfo {author} {\bibfnamefont {G.~T.}\ \bibnamefont
  {Silva}},\ }\bibfield  {title} {\bibinfo {title} {Off-axis scattering of an
  ultrasound bessel beam by a sphere},\ }\href@noop {} {\bibfield  {journal}
  {\bibinfo  {journal} {IEEE transactions on ultrasonics, ferroelectrics, and
  frequency control}\ }\textbf {\bibinfo {volume} {58}},\ \bibinfo {pages}
  {298} (\bibinfo {year} {2011})}\BibitemShut {NoStop}%
\bibitem [{\citenamefont {Mitri}(2009)}]{mitri2009Bessel}%
  \BibitemOpen
  \bibfield  {author} {\bibinfo {author} {\bibfnamefont {F.}~\bibnamefont
  {Mitri}},\ }\bibfield  {title} {\bibinfo {title} {Acoustic radiation force of
  high-order bessel beam standing wave tweezers on a rigid sphere},\
  }\href@noop {} {\bibfield  {journal} {\bibinfo  {journal} {Ultrasonics}\
  }\textbf {\bibinfo {volume} {49}},\ \bibinfo {pages} {794} (\bibinfo {year}
  {2009})}\BibitemShut {NoStop}%
\bibitem [{\citenamefont {Mitri}(2015{\natexlab{b}})}]{mitri2015}%
  \BibitemOpen
  \bibfield  {author} {\bibinfo {author} {\bibfnamefont {F.}~\bibnamefont
  {Mitri}},\ }\bibfield  {title} {\bibinfo {title} {Acoustic radiation force on
  oblate and prolate spheroids in bessel beams},\ }\href@noop {} {\bibfield
  {journal} {\bibinfo  {journal} {Wave Motion}\ }\textbf {\bibinfo {volume}
  {57}},\ \bibinfo {pages} {231} (\bibinfo {year}
  {2015}{\natexlab{b}})}\BibitemShut {NoStop}%
\end{thebibliography}%

\end{document}